\documentclass[paper]{JHEP3}
\usepackage{amssymb,enumerate}
\usepackage{amsmath}
\usepackage{bbm}
\usepackage{url}
\usepackage{cite}
\usepackage{graphics}
\usepackage{xspace}
\usepackage{epsfig}
\usepackage{subfigure}
\newcommand\pt{p_t}
\newcommand\kt{k_t}

\newcommand\MCATNLO{{\tt MC@NLO}}
\newcommand\POWHEG{{\tt POWHEG}}
\newcommand\POWHEGBOX{{\tt POWHEG BOX}}
\newcommand\PYTHIA{{\tt PYTHIA}}

\newcommand\MCFM{{\tt MCFM}}
\newcommand\MadGraph{{\tt MadGraph}}

\def\({\left(} 
\def\){\right)} 

\def\beq{\begin{equation}}
\def\beqn{\begin{eqnarray}}
\def\eeq{\end{equation}}
\def\eeqn{\end{eqnarray}}

\title{$W^+W^-$, $WZ$ and $ZZ$ production in the \POWHEGBOX{}}
\vfill
\author{Tom Melia \\
Rudolf Peierls Centre for Theoretical Physics, 1 Keble Road, University of Oxford, UK\\
E-mail: \email{t.melia1@physics.ox.ac.uk}}

\author{Paolo Nason \\
INFN, Sezione di Milano Bicocca, Italy\\
E-mail: \email{Paolo.Nason@mib.infn.it}}

\author{Raoul R\"ontsch \\
Rudolf Peierls Centre for Theoretical Physics, 1 Keble Road, University of Oxford, UK\\
E-mail: \email{r.rontsch1@physics.ox.ac.uk}}

\author{Giulia Zanderighi \\
Rudolf Peierls Centre for Theoretical Physics, 1 Keble Road, University of Oxford, UK\\
E-mail: \email{g.zanderighi1@physics.ox.ac.uk}}


\abstract{We present an implementation of the vector boson pair
  production processes $ZZ$, $W^+W^-$ and $WZ$ within the \POWHEG{}
  framework, which is a method that allows the interfacing of NLO
  calculations to shower Monte Carlo programs.  The implementation is
  built within the \POWHEGBOX{} package. The $Z/\gamma^*$ interference,
  as well as singly resonant contributions, are properly included. We
  also considered interference terms arising from identical leptons in
  the final state. As a result, all contributions leading to the desired
  four-lepton system have been included in the calculation, with the
  sole exception of the interference between $ZZ$ and $W^+W^-$ in the
  production of a pair of same-flavour, oppositely charged fermions
  and a pair of neutrinos, which we show to be fully
  negligible. Anomalous trilinear couplings can be also set in the
  program, and we give some examples of their effect
  at the LHC.  We have made the
  relevant code available at the \POWHEGBOX{} web site.
}

\begin{document}

\section{Introduction}
The pair production of electroweak vector bosons is of great interest
to the particle physics community. The leptonic decay of electroweak
boson pairs has been intensively studied at the Tevatron
\cite{Abazov:2004kc,Abazov:2005ys,Acosta:2005mu,:2007rab,%
  Abulencia:2007tu,Aaltonen:2008mv,:2009us,Abazov:2009ys,%
  Abazov:2011td}, while first measurements of $W^+W^-$ production were
recently published by both ATLAS \cite{Aad:2011kk} and CMS
\cite{Chatrchyan:2011tz}.

Vector boson pair production is interesting for various reasons.
First of all, the process is interesting in itself, as a test of the
non-Abelian nature of the electroweak force. Any deviation from the
Standard Model tri-vector boson couplings would indicate the presence
of new physics. Electroweak boson pair production is also an
irreducible background to moderately heavy Higgs production, where the
Higgs decays into electroweak bosons. It is therefore essential that
we are able to make accurate predictions for these processes.

In general, calculations at leading-order (LO) in perturbative QCD
(pQCD) have a large dependence on the unphysical factorization and
renormalization scales. In order to limit these uncertainties, it is
necessary to extend the pQCD calculations to next-to-leading order
(NLO). Analytic formulae for the LO and NLO matrix elements for a weak
boson pair decaying to leptons, written succinctly in the helicity
formalism, were presented in ref.~\cite{Dixon:1998py}~\footnote{Although $q\bar{q} \to W^+W^-$ with decays 
was previously considered to $\mathcal{O} ( \alpha_s)$ in Ref.~\cite{Baur:1995uv}, 
the virtual correction to the spin correlations were not computed there.}. These have been
implemented in publicly available programs, such as \MCFM{}
\cite{Campbell:1999ah,Campbell:2011bn}, enabling fast computations for
these processes.

However, the presence of soft and collinear divergences in the final
state means that any computation performed at NLO can only provide
accurate predictions for inclusive quantities. On the other hand,
exclusive quantities are often of great interest when analyzing an
event - for example, any quantity including a jet variable in weak
boson pair production. In order to compute exclusive quantities
accurately, parton shower programs must be used.  Two methods exist to
interface NLO results to parton shower programs: \MCATNLO{}
\cite{Frixione:2002ik} and \POWHEG{}
\cite{Nason:2004rx,Frixione:2007vw}. In fact, $W^+W^-$-pair production
was the first process studied in the former~\cite{Frixione:2002ik},
while $ZZ$ production was the first practical demonstration of the
latter approach \cite{Nason:2006hfa}. Furthermore, $W^+W^-$,
$W^{\pm}Z$ and $ZZ$ production have been studied within the \POWHEG{}
framework in refs.  \cite{Hamilton:2010mb,Hoche:2010pf}.
Notwithstanding this, a full, public NLO implementation of electroweak
boson pair production, including leptonic decays, $Z/\gamma^*$
interference and non-resonant graphs, is still missing, and is badly
needed by the experimental collaborations in order to reliably
simulate their backgrounds away from the resonant region. The purpose
of the present work is to fill this gap by providing NLO
implementations for weak boson pair production that include all
diagrams that lead to the desired 4-lepton final states.
The only effect not included is the interference between the $W^+W^-$
and $Z Z$ pair production when the final state consists of two
opposite charged leptons and a neutrino-antineutrino pair of the same
family.  We show, however, that this effect is fully negligible.

A formally NNLO contribution to these processes, initiated by gluon
fusion, is also known to be important.  We do not include it in our
program. Calculations of these processes are already available in the
literature~\cite{Binoth:2008pr,Binoth:2005ua,Binoth:2006mf,Campbell:2011bn},
and in particular the gg2ZZ and gg2WW generators of
refs.~\cite{Binoth:2008pr,Binoth:2005ua,Binoth:2006mf} can be easily interfaced
to shower program, and are currently used by the experimental collaborations
for this purpose.

The rest of this paper is organized as follows. In section
\ref{sec:implementation}, we describe the NLO calculation of the
vector boson pair production processes, and the implementation of
these processes in the \POWHEGBOX{} framework \cite{Alioli:2010xd}.
We present our results in
section \ref{sec:results}. The effects of the \POWHEG{} implementation
and subsequent parton showering, common to all three vector boson pair
production processes, are demonstrated for $ZZ$ production. We also
discuss the effects of anomalous trilinear boson couplings in $W^+W^-$
and $WZ$ production, focusing on the potential of the LHC to improve on the
existing bounds on these couplings. In section \ref{sec:conclu} we
give our conclusions. In the two Appendices we discuss some technical
details. 

\section{Description of the implementation}
\label{sec:implementation}
We took the matrix elements for the Born, real and virtual amplitudes
from the calculations of Ref.~\cite{Dixon:1998py}, borrowing heavily
from their implementation in the \MCFM{} package
\cite{Campbell:1999ah,Campbell:2011bn}.  At variance with the \MCFM{}
package, however, we have used the Breit-Wigner form for the off-shell
vector bosons propagators
\newcommand\sV{{\scriptscriptstyle V}}
\begin{equation}
\frac{1}{s_\sV - M_\sV^2 + i\Gamma_\sV M_\sV},
\end{equation}
rather than the Baur-Zeppenfeld form \cite{Baur:1995aa}.
We can also enforce the use of the running width variant (see
manuals for details) 
\begin{equation}
\frac{1}{s_\sV - M_\sV^2 + i\Gamma_\sV s_\sV/M_\sV}.
\end{equation}
We have seen, however, no appreciable differences when this last form is
used.

$Z/\gamma^*$ interference effects are included.
Singly resonant amplitudes, which are essential if we wish to describe
kinematic regions where only one of the vector bosons is resonant
\cite{Campbell:2011bn}, are also included.
\FIGURE[t]{
\includegraphics[angle=0,scale=0.78]{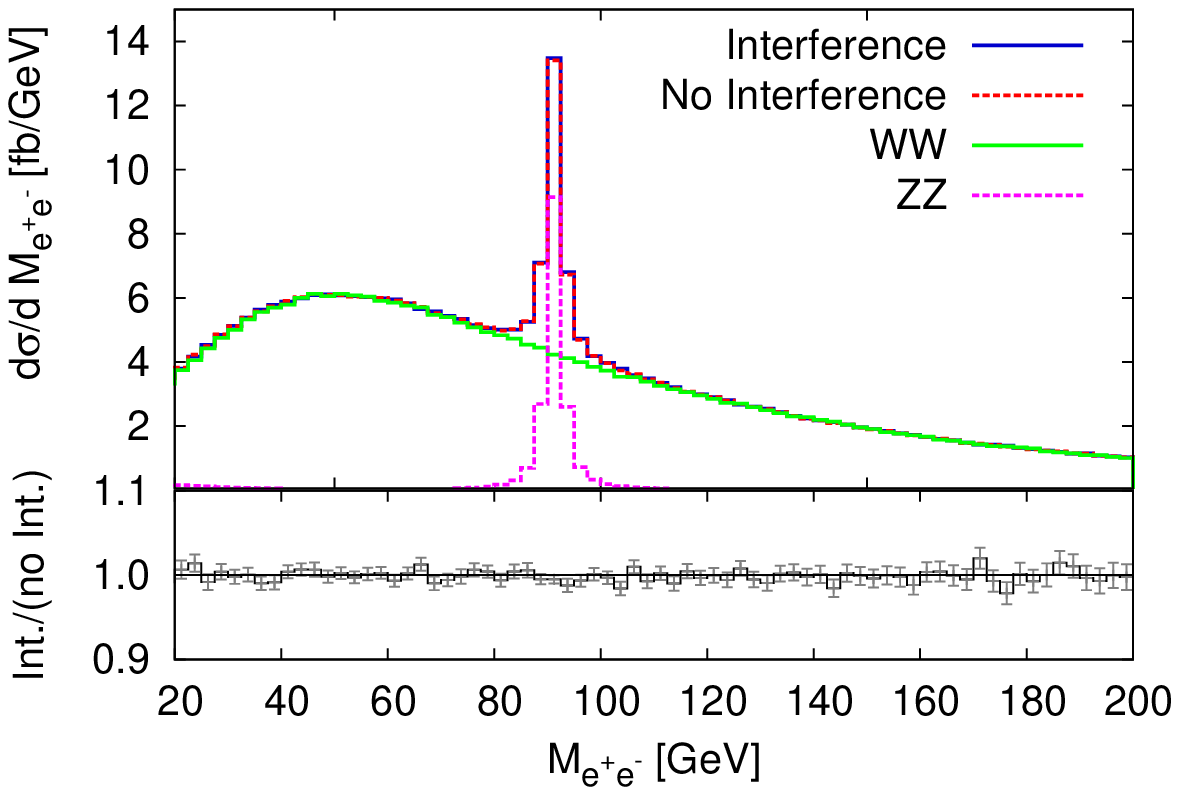}
\caption{Distribution of $M_{e^+e^-}$ in the process $pp \to e^-e^+
  \nu_e \bar{\nu}_e$ at leading order and centre of mass energy of 7
  TeV. This demonstrates the negligible effect of interference between
  $W^+W^-$ and $ZZ$ production of this final state. The solid green and
  magenta lines show the separate contribution from $W^+W^-$ and $ZZ$
  production respectively.}
\label{fig:WWZZ}}
Furthermore, we include interference terms when
there are identical leptons in the final state, as in $ZZ\to e^- e^+
e^- e^+$, and $ZW^{\pm} \to e^- e^+ e^{\pm} \nu$. These effects are
small, but become relevant when all oppositely charged lepton pairs
are far off-resonance.\footnote{Since interference effects are very
  small, and their calculation more than doubles the computing time,
  it is possible to switch them off by setting an appropriate flag
  in the {\sf powheg.input} file. Similarly, it is possible to generate
  on-shell
  vector bosons in the zero-width approximation or to remove the
  single resonant diagrams, as illustrated in the implementation
  manuals.}  Interference terms of this sort never arise in
$W^+W^-$-production. However, $W^{+}W^{-}$ and $ZZ$ production may lead to
identical final states, such as $e^-e^+ \nu_e \bar{\nu}_e$.
We have studied the corresponding interference effect at leading order with
Madgraph~\cite{Alwall:2007st}. As shown for example in
Fig.~\ref{fig:WWZZ} in the case of the $M_{e^+e^-}$ distribution, such
effects are completely negligible, and are therefore not included in
the following.  Thus, if one wishes to investigate leptonic final
states like $e^-e^+ \nu_e \bar{\nu}_e$, both a $W^+W^-$ and a $ZZ$ sample
should be generated separately.

The construction of the Born phase space requires some care, in
order to efficiently probe the resonance regions. In Appendix~A
we discuss this problem with some detail. Furthermore, as in the
case of $W$ production, a problem due to vanishing Born configurations
has to be dealt with. This is explained in Appendix~B.

The three programs perform quite differently. The $WZ$ program
can complete the preparation stage in roughly one hour, involving
about 6 million calls, and can generate one million events (at the Les
Houches level) in six hours.  The $WW$ program has similar
performance, but the preparation stage may require twice as much. The
$ZZ$ program is much slower. Even without including interference for
identical fermions, the preparation stage takes of the order of 6
hours, and one million events are generated in 40 hours. It is thus
convenient, in this case, to use the \POWHEGBOX{} facilities to run
the program in parallel.  Requiring interference costs more than a
factor of two in execution time. The poorer perfomance of the $ZZ$
program is easily tracked back the the presence of a larger number
of helicity configurations that contribute to the process.

Several checks have been performed to test the correctness of the
implementation. The Born and real
matrix elements have been checked against \MadGraph{}
\cite{Alwall:2007st} for an arbitrary phase-space point.  The ratio of
the residues of the single and double poles of the virtual amplitude
to the Born amplitude were compared against the analytically known
expressions. The finite part of the virtual amplitude was checked
against an independent Feynman diagram based program, developed 
by some of the authors, which uses the OPP subtraction
method \cite{Ossola:2008xq}.
The \MCFM{} program implements the same processes,
but does not include the interference terms for identical fermions.
We have thus verified that, when the interference terms are not
included or not present, the NLO output of the \POWHEG{}
implementation, which can be obtained by setting an appropriate option
in the program, is consistent within errors to that of \MCFM{} for a
large number of distributions.  Furthermore, the \POWHEG{} output at
the level of the Les Houches Interface events has been compared
for on shell $Z$ bosons to the
previous implementation of
ref.~\cite{Nason:2006hfa}. Full
agreement was found, thus demonstrating that the mechanism used by
\POWHEG{} to generate radiation operates in the same way as in the
implementation of ref.~\cite{Nason:2006hfa}, which in turn was
thoroughly compared to the \MCATNLO{} implementation.
\newcommand\sZ{{\scriptscriptstyle Z}}
\newcommand\sW{{\scriptscriptstyle W}}
\section{Results}
\label{sec:results}
We now present some results obtained with our \POWHEG{}
implementation. Our basic setup for the electroweak input parameters
is the following. We use $M_\sZ$, $M_\sW$ and $G_\mu$ as basic parameters
of the Standard Model (SM), and define all remaining quantities according
to the leading order SM relations. Thus we have
\begin{equation}
\cos\theta_w=\frac{M_\sW}{M_\sZ},\quad
\alpha_{\rm em}=\frac{\sqrt{2}G_\mu M_\sW^2}{\pi} \sin^2 \theta_w
\end{equation}
This corresponds to the so called ``$G_\mu$ scheme'', advocated in
ref.~\cite{Dittmaier:2001ay}. We adopt the following
PDG~\cite{Nakamura:2010zzi} values for the independent parameters
\begin{equation}
M_\sZ=91.1876\,{\rm GeV},\quad M_\sW=80.399\,{\rm GeV},\quad
G_\mu=1.166364\times 10^{-5}\,{\rm GeV}^{-2}\;.
\end{equation}
\newcommand\sS{{\scriptscriptstyle S}}
For the width of the $W$ and $Z$ bosons, we take their LO value
computed in the above scheme, except that we correct the hadronic
width with a factor $(1+\alpha_\sS(M_\sW)/\pi)$ and
$(1+\alpha_\sS(M_\sZ)/\pi)$, respectively.  In this way, the branching
fractions of the $W/Z$ consistently add up to one. For the $W$ and $Z$
width we get $2.0997$~GeV and $2.5096$~GeV, respectively. These values
are around half a percent away from the current PDG
values~\cite{Nakamura:2010zzi}, $2.085 \pm 0.042$~GeV and $2.4952 \pm
0.0023$~GeV, respectively, so that, in fact, the measured values could
be used instead.  Details can be found in the file {\tt
  smcouplings.f}, which can easily be modified by users
preferring other ways to implement SM couplings.  As default, we set
the factorization and renormalization scales equal to the mass of the
four-lepton system, $M_{4l}$, and our default PDF set is the MSTW 2008
NLO \cite{Martin:2009iq}.

The three processes we examine are very similar to one another. For
each one of them we have analyzed several distributions, such as: the
rapidity, pseudorapidity, transverse momentum and invariant mass of
each lepton, each combination of lepton-lepton and lepton-antilepton pairs,
and the four lepton system; the distance in rapidity, pseudorapidity,
azimuth and $\Delta R$ distance (where
$\Delta R=\sqrt{\Delta \phi^2+\delta\eta^2}$) of all lepton-antilepton
pairs; and the rapidity distribution of the hardest jet and its
distance in rapidity from the four lepton system, with several transverse
momentum cuts. For each distribution, the pure NLO result has been
compared to the \POWHEG{} result before the shower. All distributions
behave as expected. In the present paper we illustrate a selected
set of distributions with realistic cuts in the $ZZ$ case only.
We compare the various levels of approximation that one
can include in the NLO calculation, and
the full NLO result with the event
output by \POWHEG{} before and after
the shower.

We remind the reader
that the \POWHEG{} output can be easily interfaced to all general
parton shower event
generators~\cite{Sjostrand:2006za,Corcella:2000bw,Bahr:2008pv,Sjostrand:2007gs} that comply with the Les Houches event files format~\cite{Alwall:2006yp}.
In the present work we use \PYTHIA{} version 6.4.25 \cite{Sjostrand:2006za}.

We illustrate our implementation of $W^+W^-$ and $WZ$ production by
studying the effect that anomalous couplings have on these processes
at the LHC.

\subsection{$ZZ$ Production}
We first illustrate the differences that arise at different levels of
approximation in our generator
in the $ZZ$ case. In Fig.~\ref{fig:approx} \FIGURE[t]{
\includegraphics[angle=0,scale=1]{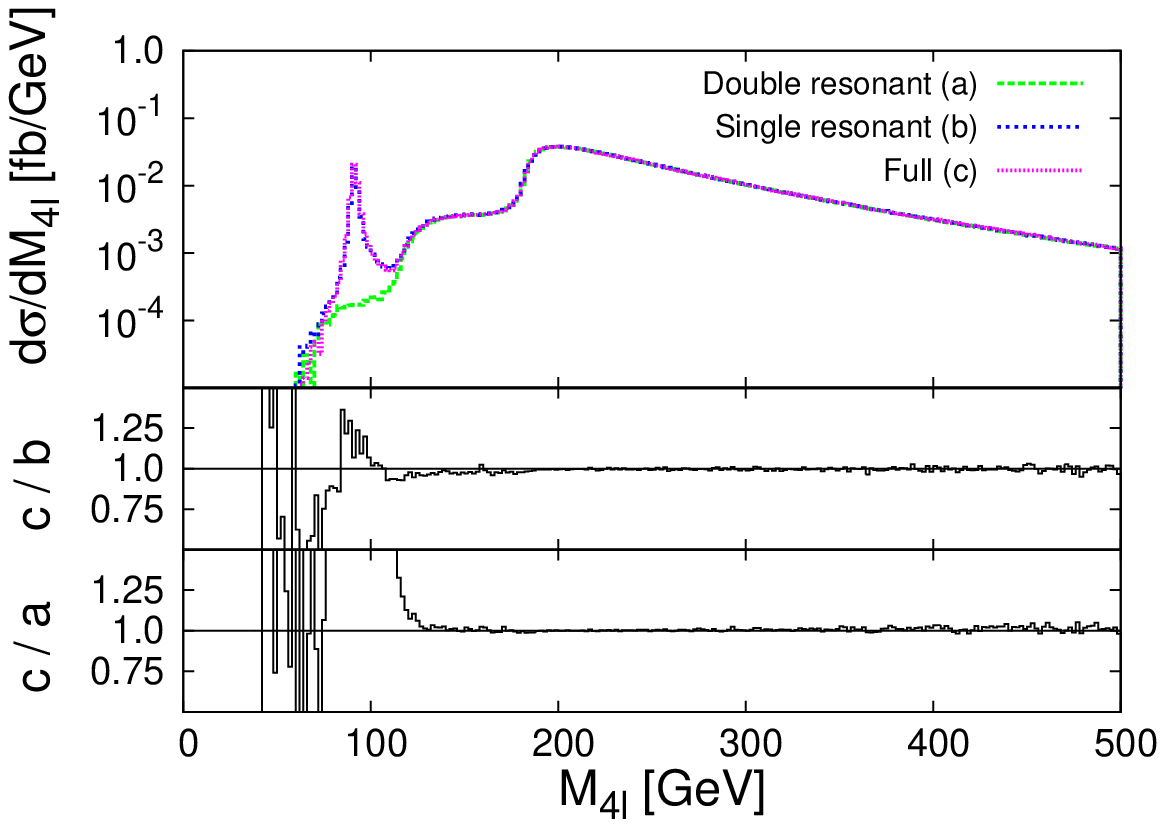}
\caption{The effect of different levels of approximation of the calculation of $pp\to e^+e^- e^+e^-$ at 7 TeV for the invariant mass of the four-lepton system.}
\label{fig:approx}}
we show three predictions for the invariant mass of the four-lepton system
in $ZZ$ production, when both $Z$ bosons decay into electrons. The
cuts are as documented further on, but for the purpose of
understanding the present plot, the only relevant cut is the
$20$ GeV one on the invariant mass of opposite sign leptons.  The curve
labelled as (c) is the full calculation of the four lepton production
process, and thus is the more accurate result that our generator can
provide. In (b), we suppress the interference effects due to identical
fermion pairs in the final state. The ratio of (c) over (b) is
presented in the lower panel. We see that there are sensible
differences around the $Z$ peak. This is easily understood. The
singly-resonant production mechanism includes the production of a
single $Z$ boson that decays into four leptons, giving rise to the $Z$
peak observed in the figure.  In this decay process both lepton pairs
are off resonance, and thus we can expect non-negligible interference
effects. The knee visible in the figure at roughly $M_{4l}\approx$
110-120~GeV is due to the onset of the production of an on-shell $Z$
decaying into a lepton pair, with the second lepton pair arising from
a $Z^*/\gamma^*$ decay (this mechanism is displaced to
110-120~GeV because of the 20 GeV cut on the minimum invariant mass
for a lepton pair).  In this kinematic regime, with an on-shell $Z$
decaying into a lepton pair, the interference is suppressed by the
phase space.  Finally, the curve labelled (a) does not include single
resonant graphs or interference for identical fermions. It is in good
agreement with the full result only if we are sufficiently beyond the
threshold for the production of one on-shell $Z$ boson.  The
approximation (b) is presently implemented in the \MCFM{} package.
The approximation (a) is implemented, for example, in \PYTHIA{}, where
off-shell effects and $Z/\gamma^*$ interference are accounted for, but
singly resonant graphs are not included.
Because of the small differences between the single resonant (b) and
the full calculation (c), for simplicity we proceed by considering
decays to $e^+e^- \mu^+\mu^-$ only.  

\newcommand\sR{{\scriptscriptstyle R}}
\newcommand\sF{{\scriptscriptstyle F}}
\begin{table} 
\begin{center}
\begin{tabular}{|c|c|c|c|}
\hline
        & MSTW2008                 & CT10                    & NNPDF2.1 \\
\hline
LO (fb) &$14.61(1)^{+0.19}_{-0.31}$ & $14.44(1)^{+0.19}_{-0.31}$&$14.61(1)^{+0.21}_{-0.32}$ \\
\hline
NLO (fb)&$18.24(1)^{+0.37}_{-0.31}$ & $17.95(1)^{+0.35}_{-0.29}$&$18.21(1)^{+0.37}_{-0.30}$ \\
\hline
\end{tabular}
\end{center}
\caption{Total cross section for $pp \to ZZ \to e^+ e^- \mu^+ \mu^-$,
  with the only cut $M_{l^+l^-}>20$ GeV on both lepton pairs. The
  central values are for the scales $\mu_\sF = \mu_\sR = M_{4l}$, and
  the integration error in the final digit is shown in
  parentheses. The theoretical uncertainty is obtained by doubling
  and halving independently the scales $\mu_\sF$ and $\mu_\sR$ with respect
  to their central value, excluding
  the combinations that yield $\mu_\sF/\mu_\sR=4$ and $\mu_\sR/\mu_\sF=4$.}
\label{tab:xs}
\end{table}
In the following, we employ a cut of $M_{l^+l^-}> 20$ GeV for
each electrically charged, same-flavour lepton-antilepton pair
$l^+l^-$, which ensures that divergences due to low-virtuality photon emissions
are avoided.
We remark here that, if interference effects due to identical fermions
are neglected, the same flavour and different flavour cross sections
can be simply related; for example, the $e^+e^-\mu^+\mu^-$ cross
section is twice the $e^+e^-e^+e^-$ one. We should however remember that
cuts applied to the final state are generally different: in the
case of $e^+e^-e^+e^-$ we limit the invariant mass of all four
possible choices of oppositely charged lepton pairs, whereas in the
$e^+e^-\mu^+\mu^-$ case there are only two such pairs. Thus,
the cross sections are related by a factor of 2 (if we neglect
interference) only if we impose additional mass cuts on the $e^+\mu^-$ and
$e^-\mu^+$ pairs in the $e^+e^-\mu^+\mu^-$ case.

In Table \ref{tab:xs}, we show the $ZZ \to e^-e^+ \mu^- \mu^+$
cross-section, at leading and next-to-leading order, and show
the effects of varying independently the
renormalization and factorization scale by
a factor of two in either direction, excluding the values that lead
to a ratio of the two scales equal to 4 or 1/4. We thus consider the
seven combinations: $\mu_\sF=\mu_\sR=M_{4l}$,  $2\mu_\sF=\mu_\sR=M_{4l}$,  
$\mu_\sF/2=\mu_\sR=M_{4l}$,  $\mu_\sF=2\mu_\sR=M_{4l}$,
$\mu_\sF=\mu_\sR/2=M_{4l}$,  $2\mu_\sF=2\mu_\sR=M_{4l}$,
$\mu_\sF/2=\mu_\sR/2=M_{4l}$, with the first one taken as the central value.
This is done for three PDF sets: MSTW08 \cite{Martin:2009iq}, CT10
\cite{Lai:2010vv} and NNPDF2.1 \cite{Ball:2011mu}. Note that we employ
NLO parton distribution functions also for LO cross-sections. The
scale uncertainty is just 1-2\% at leading order, and slightly smaller
at next-to-leading order. We observe, however, that the NLO result
does not lie within the LO scale variation band. The anomalously small
scale dependence of the LO cross-section is due to the fact that the
Born level process is of order zero in the strong coupling constant,
and thus the only scale dependence comes from the PDFs. Furthermore,
in $pp$ collisions new quark-gluon channels open up in the NLO
cross-section that compete sensibly with the LO one. Notice also that
the $gg$-initiated contribution, which is formally of NNLO (and thus
is not included in the table), is larger than the scale variation seen
here. All these considerations lead to the conclusion that scale variation
alone underestimates higher order effects for this process.
The difference in
cross-sections between different PDF sets is similar in size
to the scale variation.

\FIGURE[t]{
\includegraphics[width=0.48\textwidth]{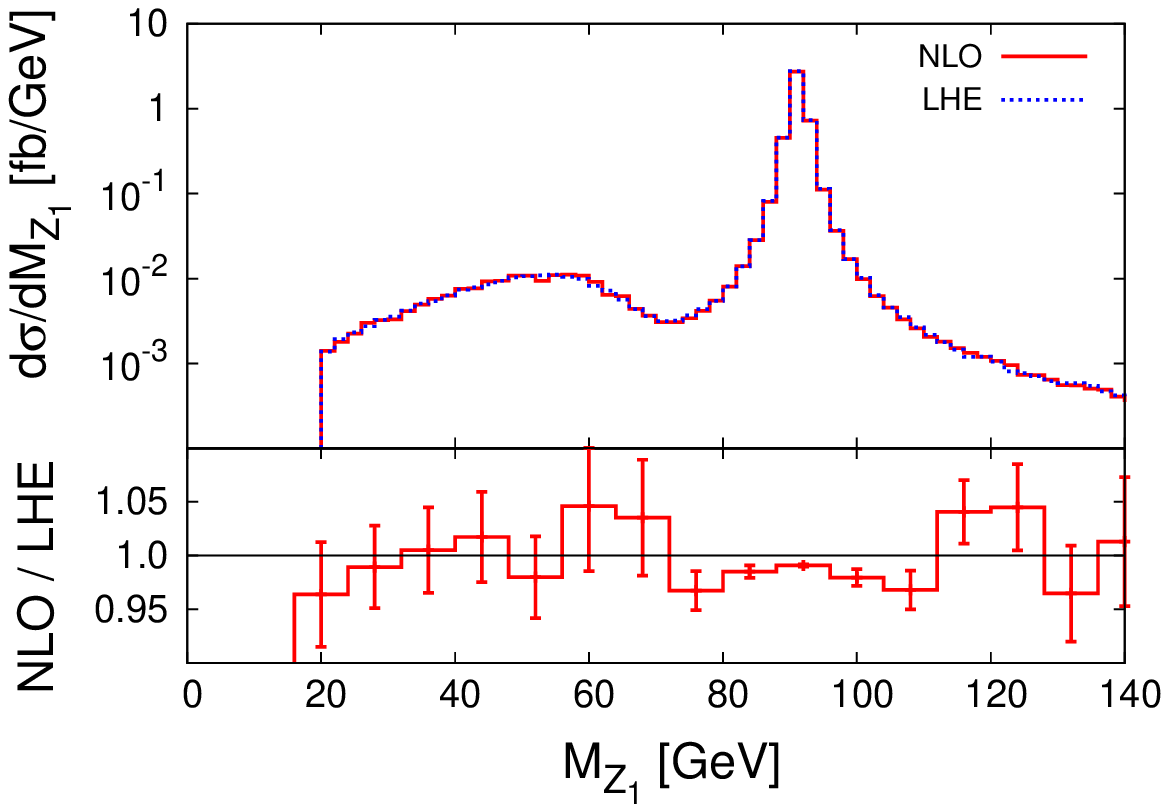}
\includegraphics[width=0.48\textwidth]{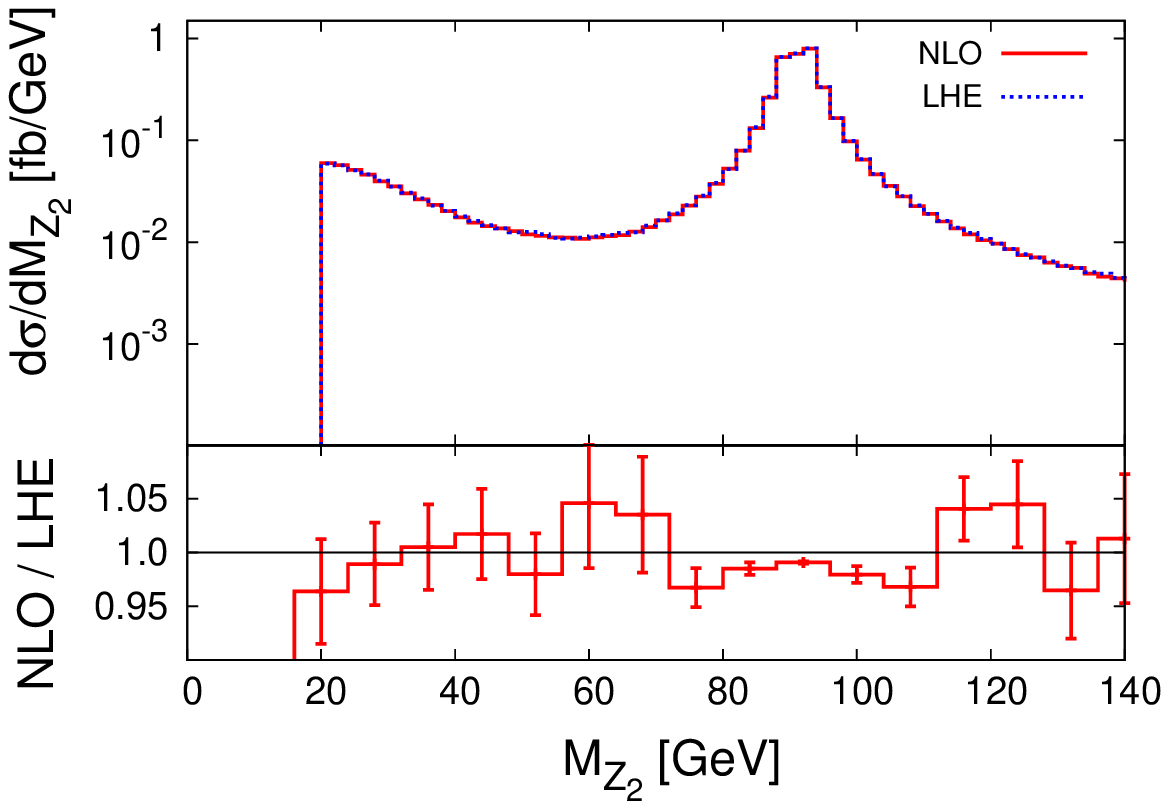}
\includegraphics[width=0.48\textwidth]{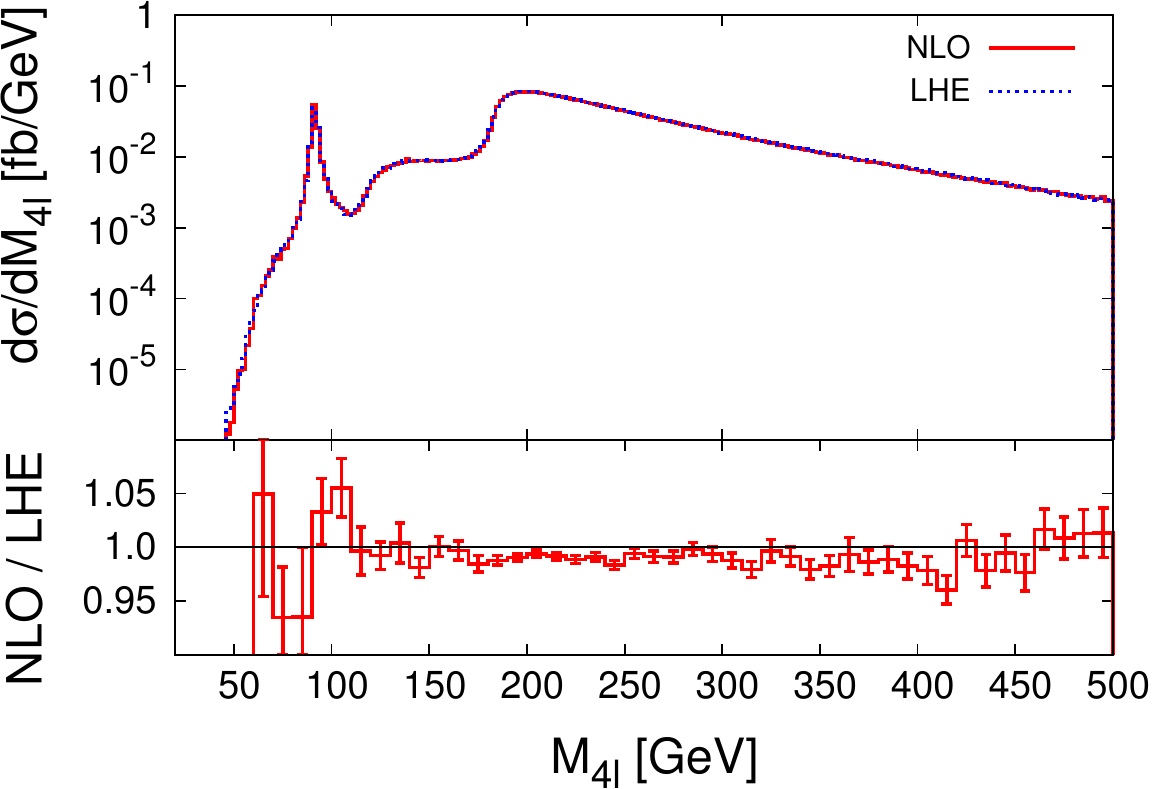}
\includegraphics[width=0.48\textwidth]{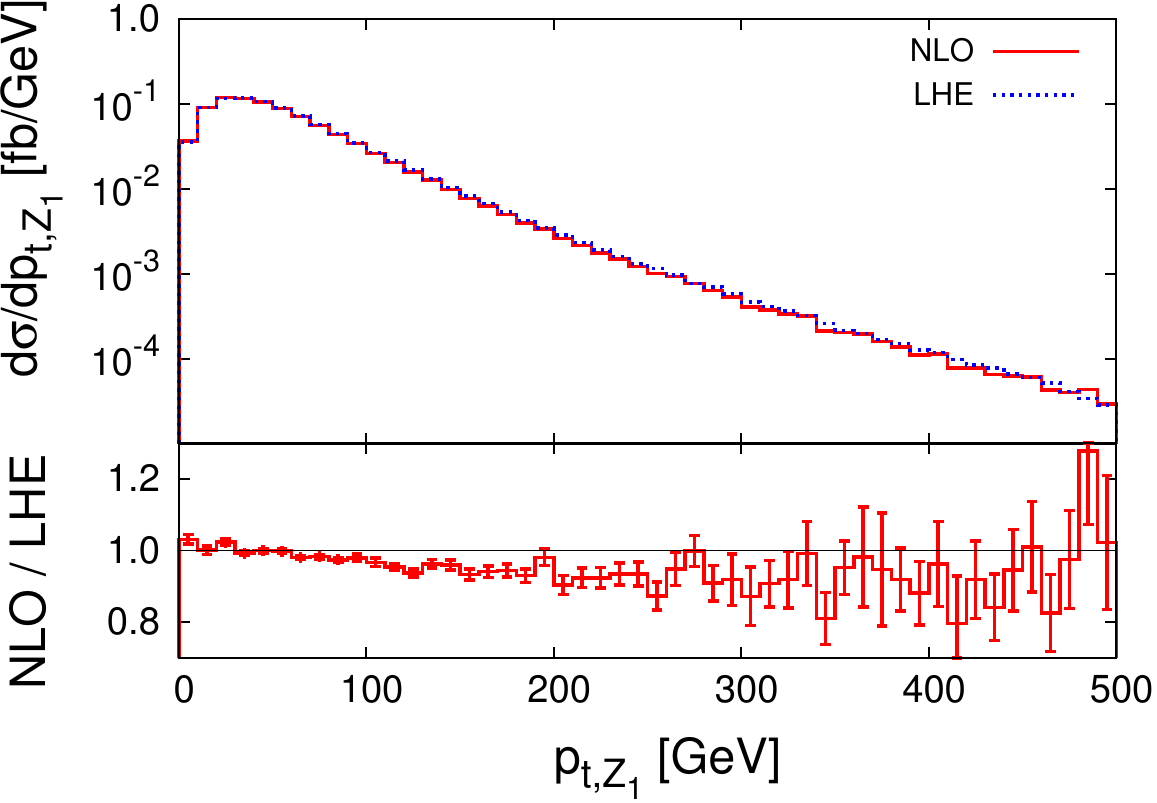}
\includegraphics[width=0.48\textwidth]{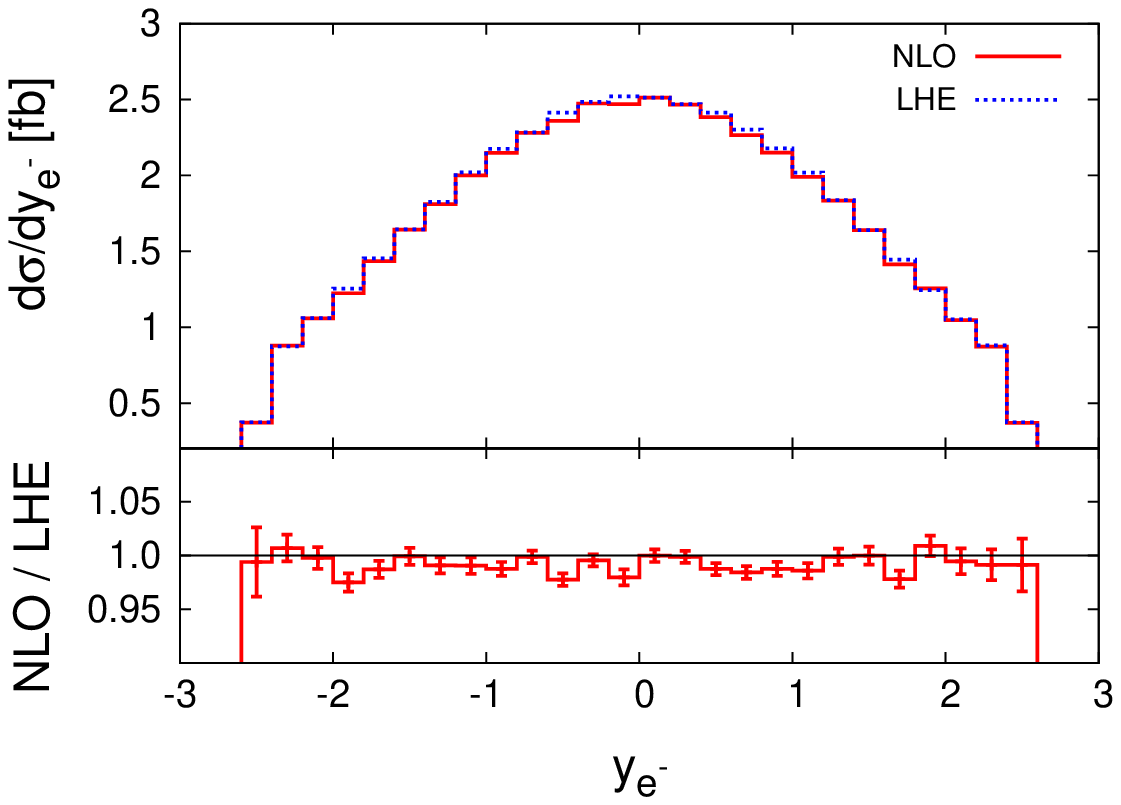}
\includegraphics[width=0.48\textwidth]{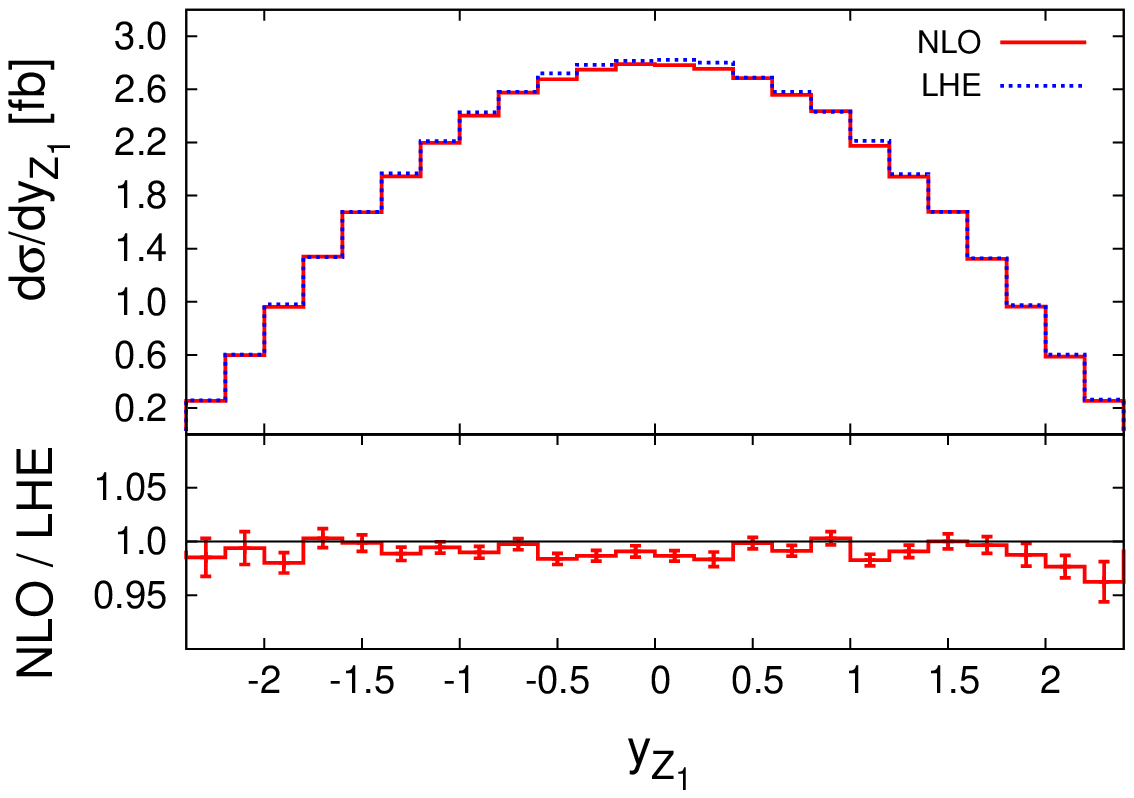}
\caption{Comparison of NLO to LHE events for $pp\to e^+e^- \mu^+\mu^-$
  at 7 TeV for a number of observables, for our default set of
  cuts. The definitions of $Z_1$ and $Z_2$ are given in the text.}
\label{fig:NLOvsLH}
}

Experience with the \POWHEG{} implementation of other processes has
taught us that, for processes that involve only initial state
radiation (ISR), one expects the effect of showers to be modest, a
fact that we will verify also in the present context.

We begin by carrying out a detailed comparison of the \POWHEG{}
results at the level of the Les Houches events output (the LHE level
from now on) with the NLO results.  We impose the following
cuts: the highest $\pt$ lepton must have $\pt>20\;$GeV, and the other
three hardest leptons must have $\pt>10\;$GeV.  We also require for
the lepton rapidities that $\vline \eta_l \vline < 2.5$. Jets are
reconstructed using the anti-$\kt$ algorithm
\cite{Cacciari:2008gp} as implemented in FastJet
\cite{Cacciari:2005hq}, with $R=0.6$. In Fig.~\ref{fig:NLOvsLH}, we
show results obtained by analyzing the LHE files directly, without
further showering, compared to the NLO result.  Six distributions are
presented: $M_{\sZ_1}$, $M_{\sZ_2}$, $M_{4l}$, $p_{t,\sZ_1}$, $y_{e^-}$, and
$y_{\sZ_1}$, where $y$ is rapidity and $M$ denotes an invariant mass
distribution.  For every phase space point, we assign $Z_1$ to the
lepton pair of the same flavour whose invariant mass is closest to
$M_\sZ$ and label the other lepton pair as $Z_2$. This
explains the difference between the first two plots. In fact, the
lepton pair closest to the $Z$ mass tends to have a much reduced
contribution from $\gamma^*$ production.  In general, we see good
agreement, at the level of few percent, between the two predictions
for these inclusive quantities. Other inclusive variables exhibit a
similar behaviour.

We now show the transverse momentum distribution of the four lepton
system, $p_{t, 4l} = |\sum_{l} \vec p_{t,l}|$, in the left plot of
Fig.~\ref{fig:ptzz}, which at NLO is identical to the transverse
momentum distribution of the jet.  \FIGURE[t]{
\includegraphics[width=0.48\textwidth]{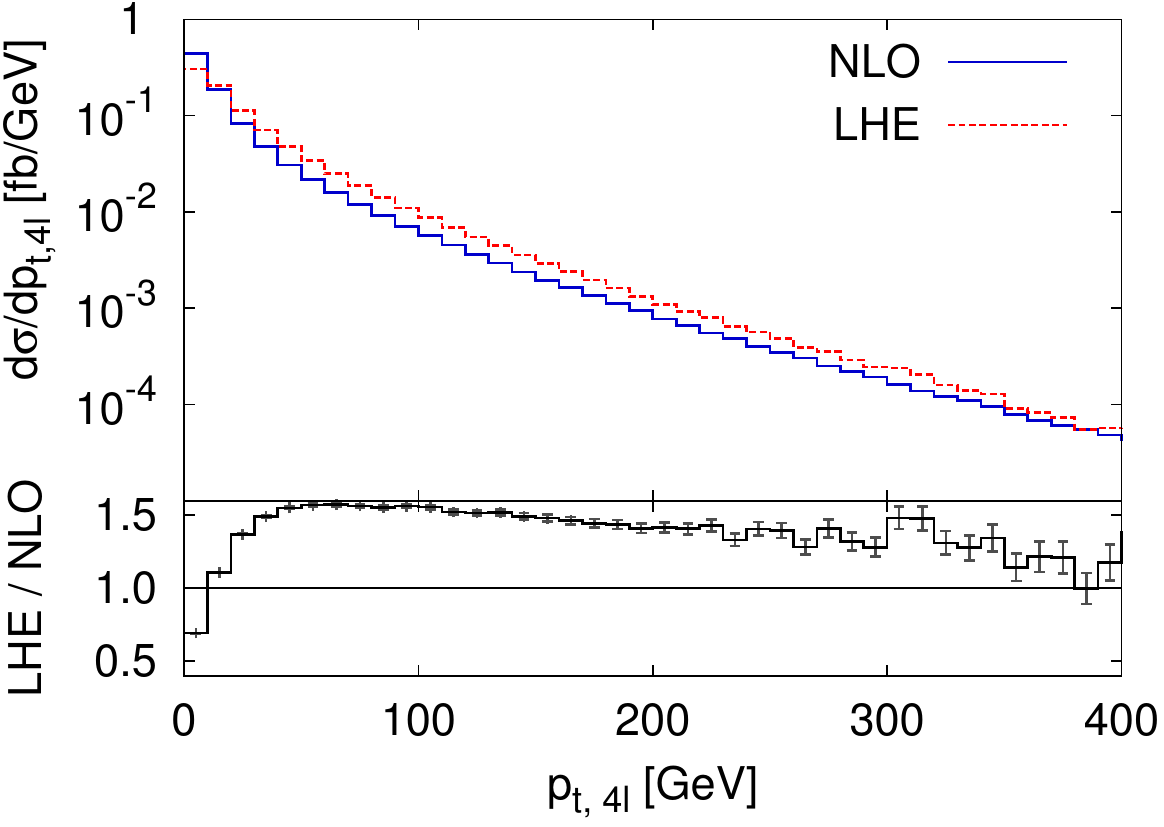}
\includegraphics[width=0.48\textwidth]{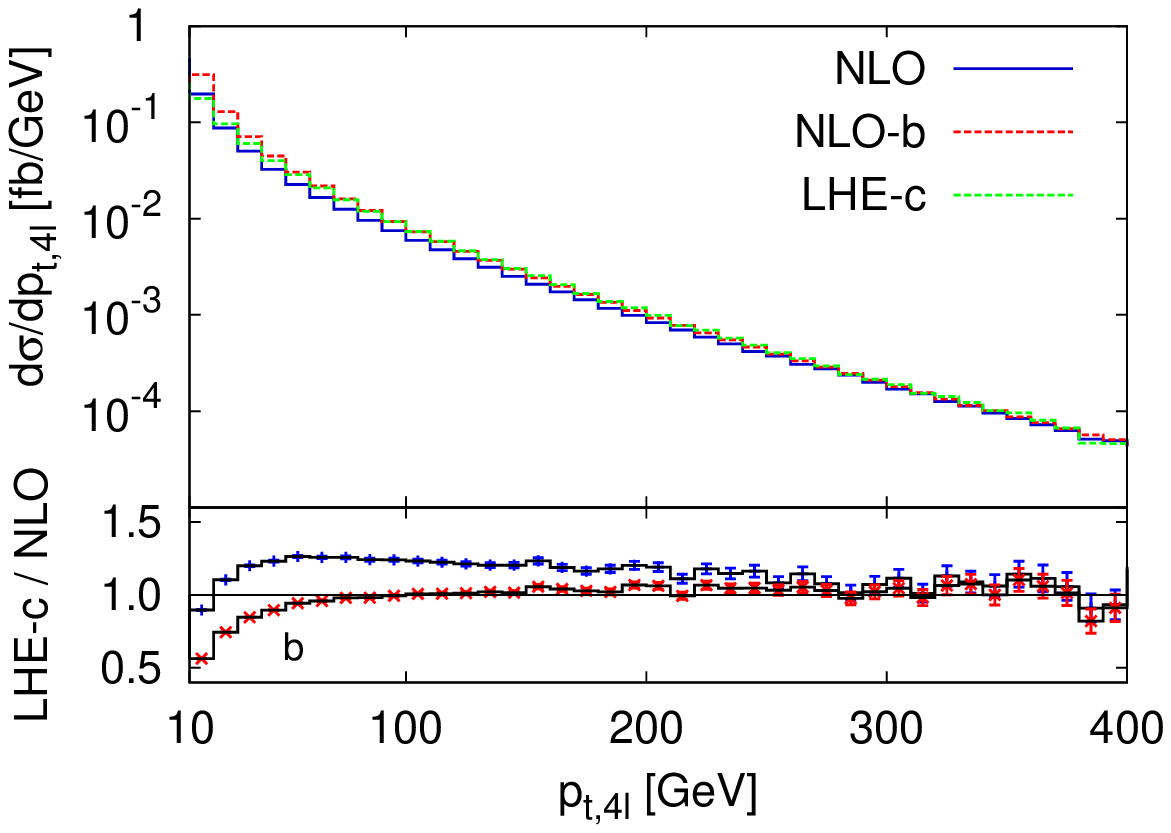}
\caption{In the left plot, we compare the transverse momentum
  distribution of the four-lepton system in the process $pp\to e^+e^-
  \mu^+\mu^-$ at 7~TeV, obtained with the fixed order NLO result, and
  at the Les Houches events level.  On the right plot the same
  comparison is carried out for the fixed order NLO result (NLO), and
  the fixed order NLO calculation with the renormalization and
  factorization scale in the real contribution set to $p_{t,4l}$
  (NLO-b), versus a \POWHEG{} result at the Les Houches event level
  with the {\tt bornonly} option set (LHE-c). In the lower panel, the
  curve labelled b is the ratio of LHE-c/NLO-b.}
\label{fig:ptzz}} We see large differences between NLO and LHE
results. At low $p_{t,4l}$, the LHE result is much smaller than the
NLO one. This fact is typically understood as originating from the
Sudakov suppression which is present in the \POWHEG{} output, but not
in the NLO result, which diverges at small $p_{t,4l}$. At large
$p_{t,4l}$ the LHE result overshoots the NLO one by about $50$\%.  An
enhancement of the large $p_{t,4l}$ tail of the \POWHEG{} distribution
is also observed in processes like single vector boson production and
gluon-fusion Higgs production. The origin of this effect is discussed
in~\cite{Alioli:2008tz,Nason:2010ap}. In essence, in \POWHEG{}, the
large transverse momentum distribution in the real emission process is
enhanced by the NLO K-factor. We thus expect an enhancement of the
order of $\sim$1.25 in our case.  However, we see that this effect
accounts only for a fraction of the enhancement observed in the
plot. The remaining enhancement is easily traced to arise from the
different choice of scales used in the NLO calculation and in the
generation of LHE events, the former being equal to the invariant mass
of the four lepton system, and the latter being instead taken equal to
its transverse momentum. These points are illustrated in the right
plot of Fig. \ref{fig:ptzz} where we plot the same quantity using the
{\tt bornonly} option in \POWHEG{}\footnote{This option eliminates the
  K-factor enhancement by replacing $\bar{B}\to B$ in the \POWHEG{}
  code, see \cite{Alioli:2008tz,Nason:2010ap} for details.}  compared
to the standard NLO result, and to the NLO result obtained by setting
the factorization and renormalization scales equal to $p_{t,4l}$ in
the real contribution. The ratio of the \POWHEG{} result with the {\tt
  bornonly} option set, over the NLO result with the $p_{t,4l}$ scale
choice, eliminates two causes of the difference (i.e. the different
scale choice, and the $\bar{B}/B$ factor), leaving only the Sudakov
effect, represented in the figure by the ratio of the LHE-c versus the
NLO-b result (curve marked as b in the lower panel). Observe that the
ratio LHE-c/NLO is about 25\% lower than the LHE/NLO ratio, thus
showing that the K-factor effect is indeed of the order of
25\%. Finally, we remark that the scale choice $\mu = p_{t,4l}$ is
more appropriate for the description of the radiation tranverse
momentum than our default one in the NLO calculation.  Thus, in this
respect \POWHEG{} does the right thing.

\FIGURE[t]{
\includegraphics[width=0.48\textwidth]{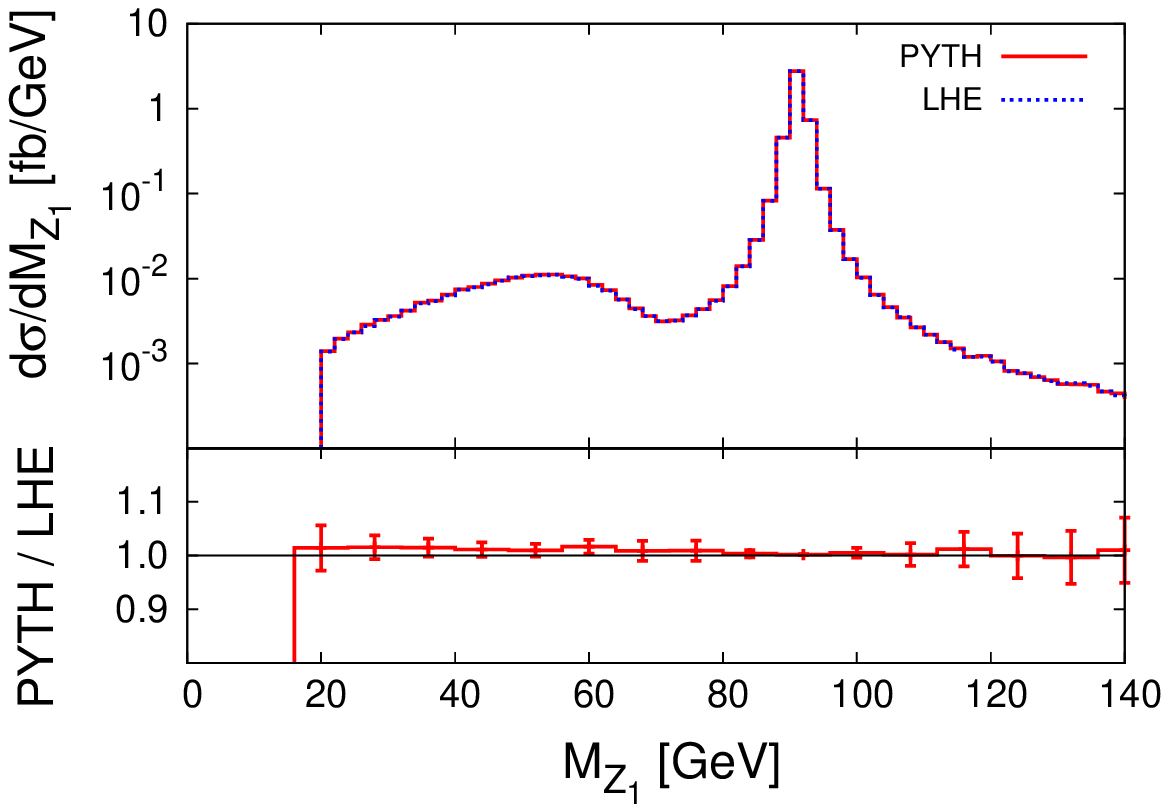}
\includegraphics[width=0.48\textwidth]{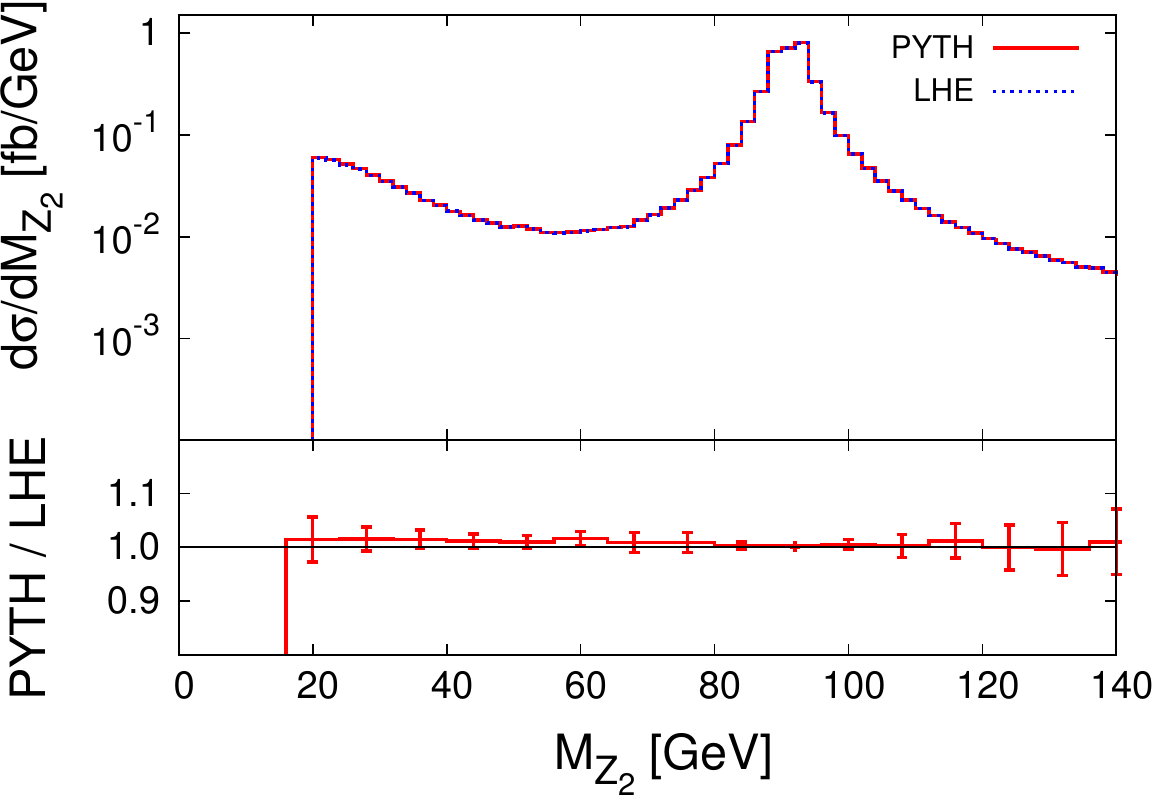}
\includegraphics[width=0.48\textwidth]{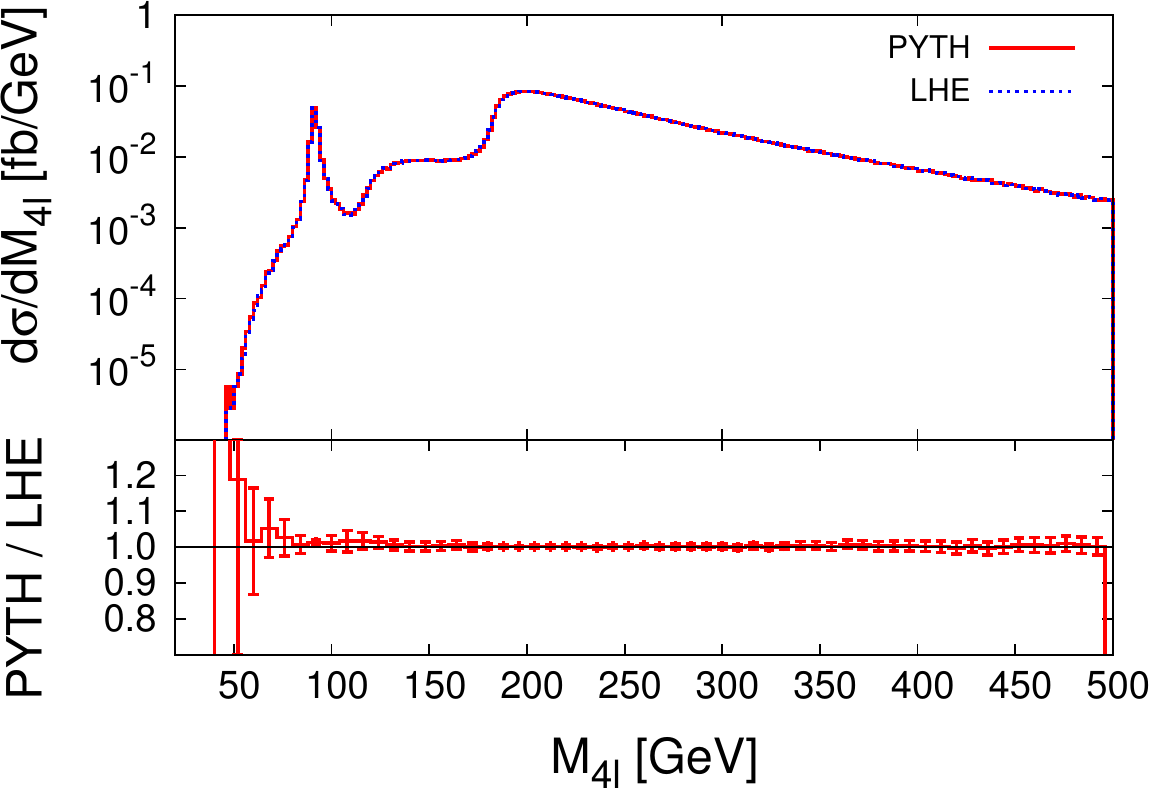}
\includegraphics[width=0.48\textwidth]{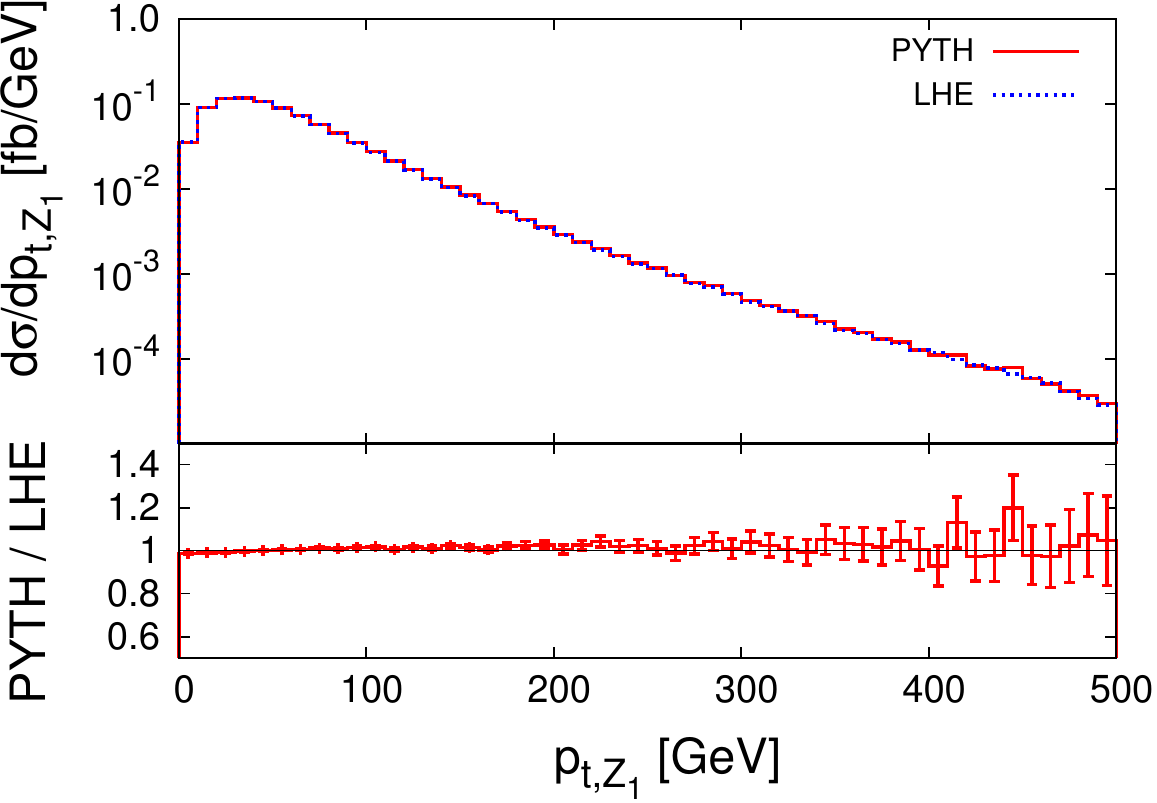}
\includegraphics[width=0.48\textwidth]{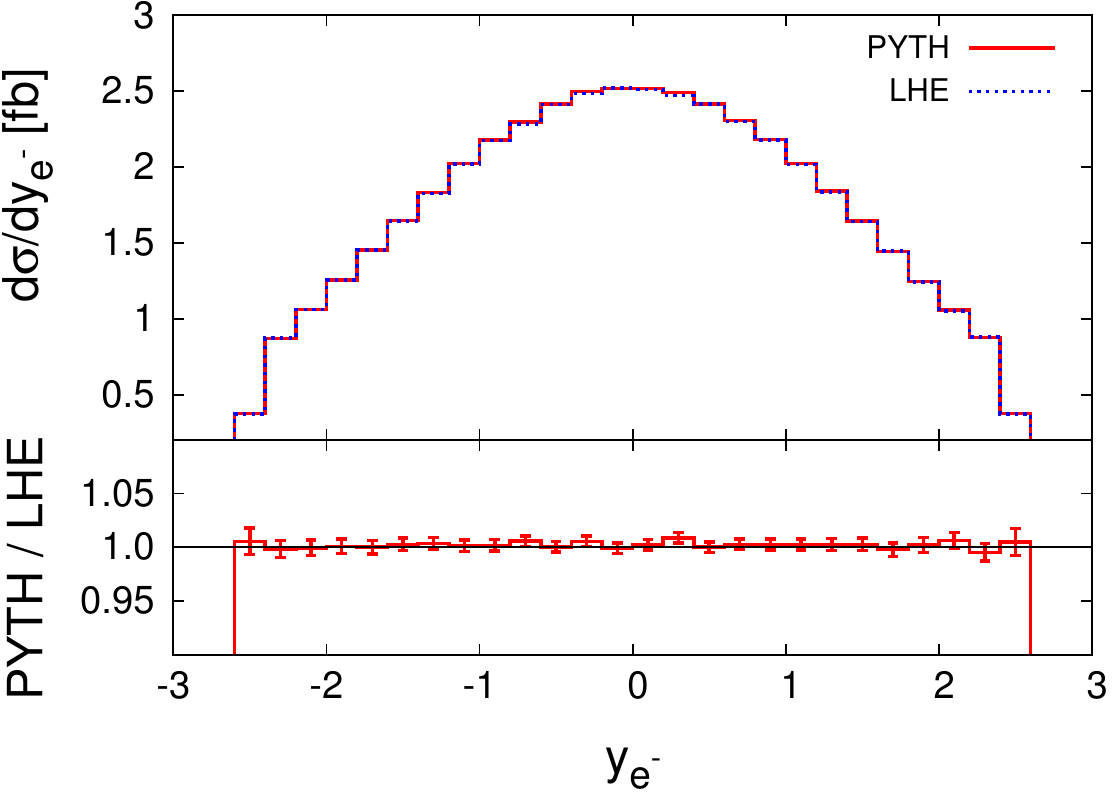}
\includegraphics[width=0.48\textwidth]{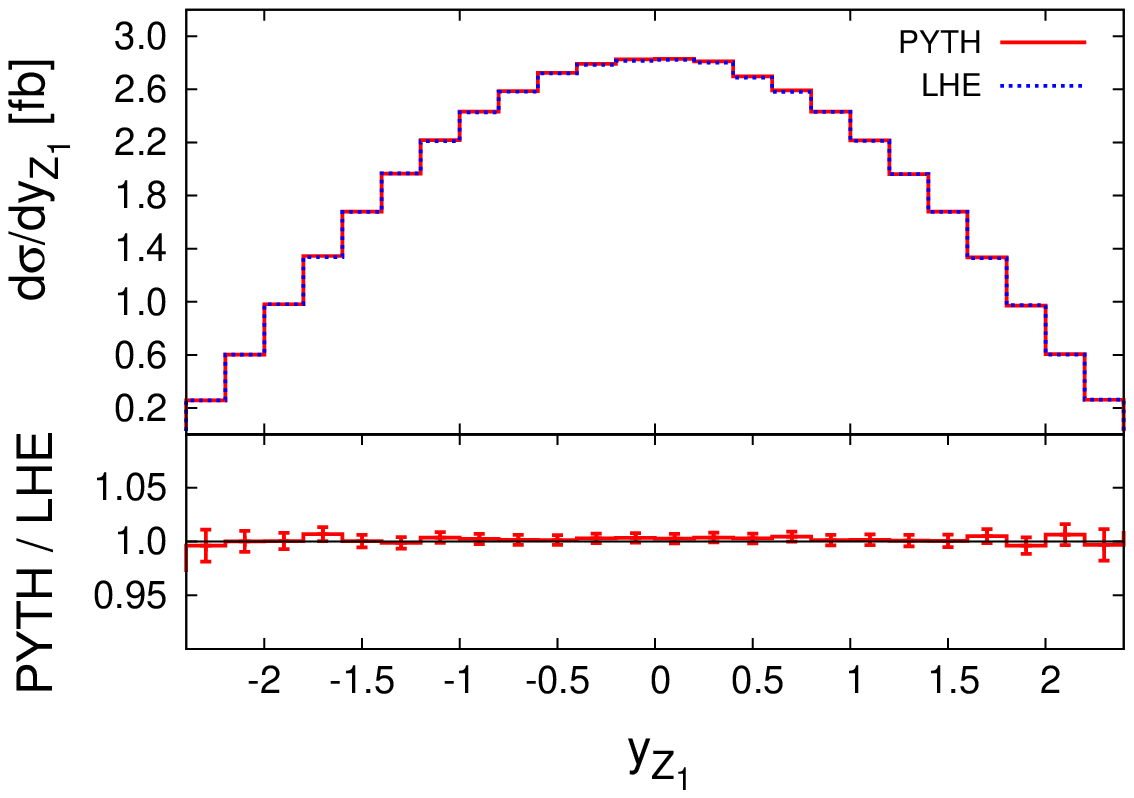}
\caption{Comparison of LHE to \PYTHIA{} showered events for $pp\to e^+e^-
  \mu^+\mu^-$ at 7 TeV for a number of observables, for our default
  set of cuts. The definitions of $Z_1$ and $Z_2$ are given in the
  text.}
\label{fig:LHEvsPYT}
} We now examine, for the same observables shown in
Fig.~\ref{fig:NLOvsLH}, the effect of a subsequent shower, obtained by
interfacing the \POWHEG{} output with \PYTHIA{}. When showering, we
turn off electromagnetic radiation.\footnote{This is adequate in this
  work for purpose of illustration, since an experiment will generally
  correct the lepton energy for electromagnetic radiation emitted
  collinearly.}
The results can be seen in Fig. \ref{fig:LHEvsPYT}. As expected, we
see very little effect from the parton shower for these
observables. We remark that in \POWHEG{}, by construction, the
invariant mass of the two- and four-lepton systems are in fact the
same at the NLO-level, LHE-level, and after parton shower (regardless
of the shower program).

\FIGURE[t]{
\includegraphics[width=0.48\textwidth]{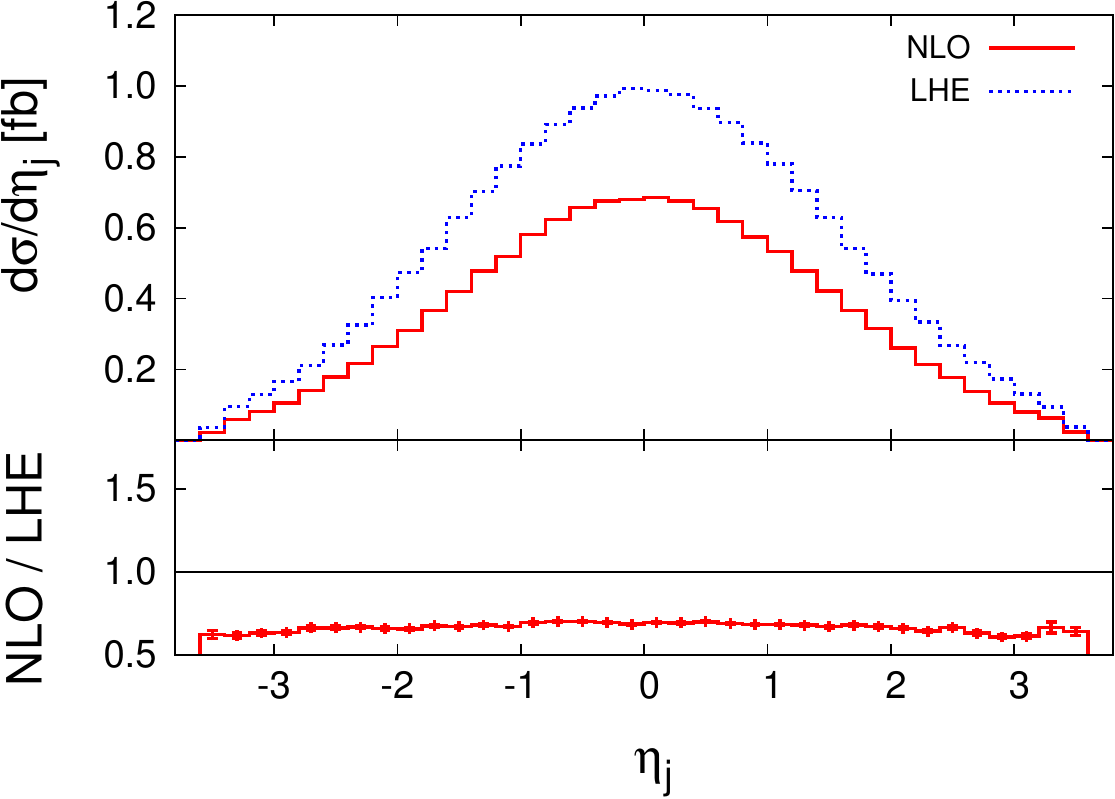} 
\includegraphics[width=0.48\textwidth]{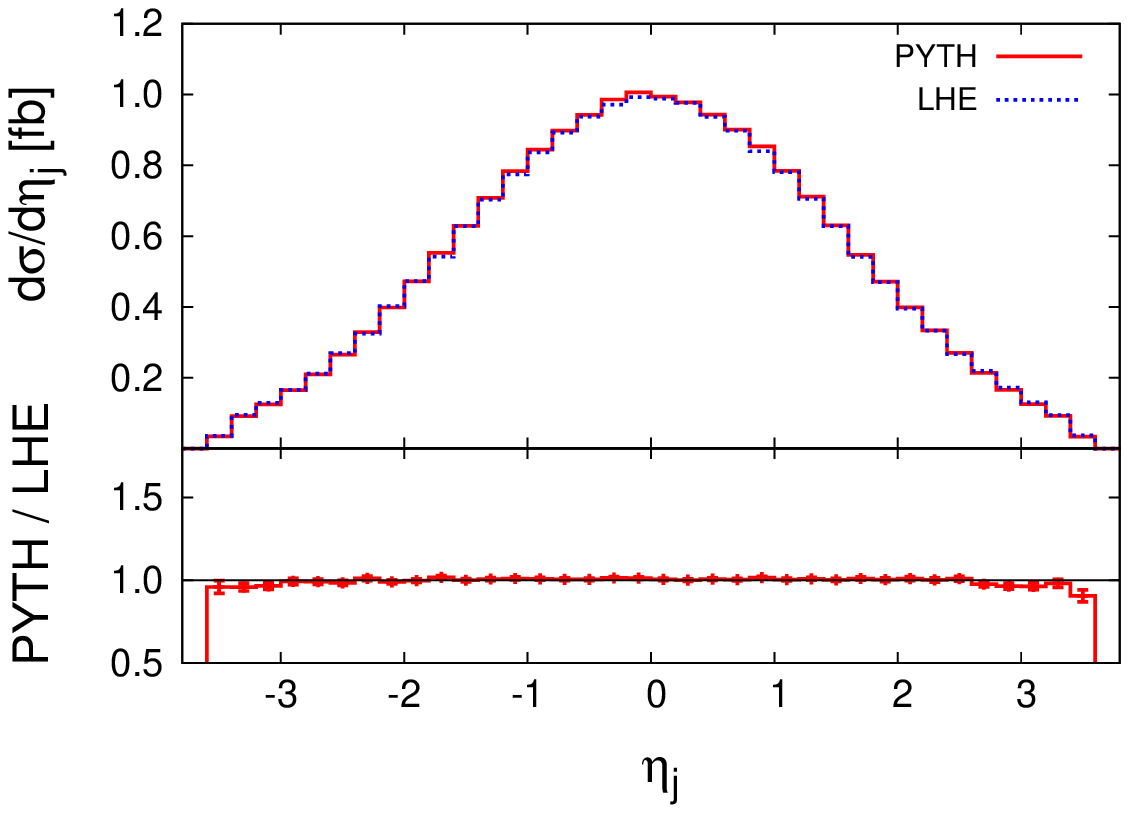} 
\includegraphics[width=0.48\textwidth]{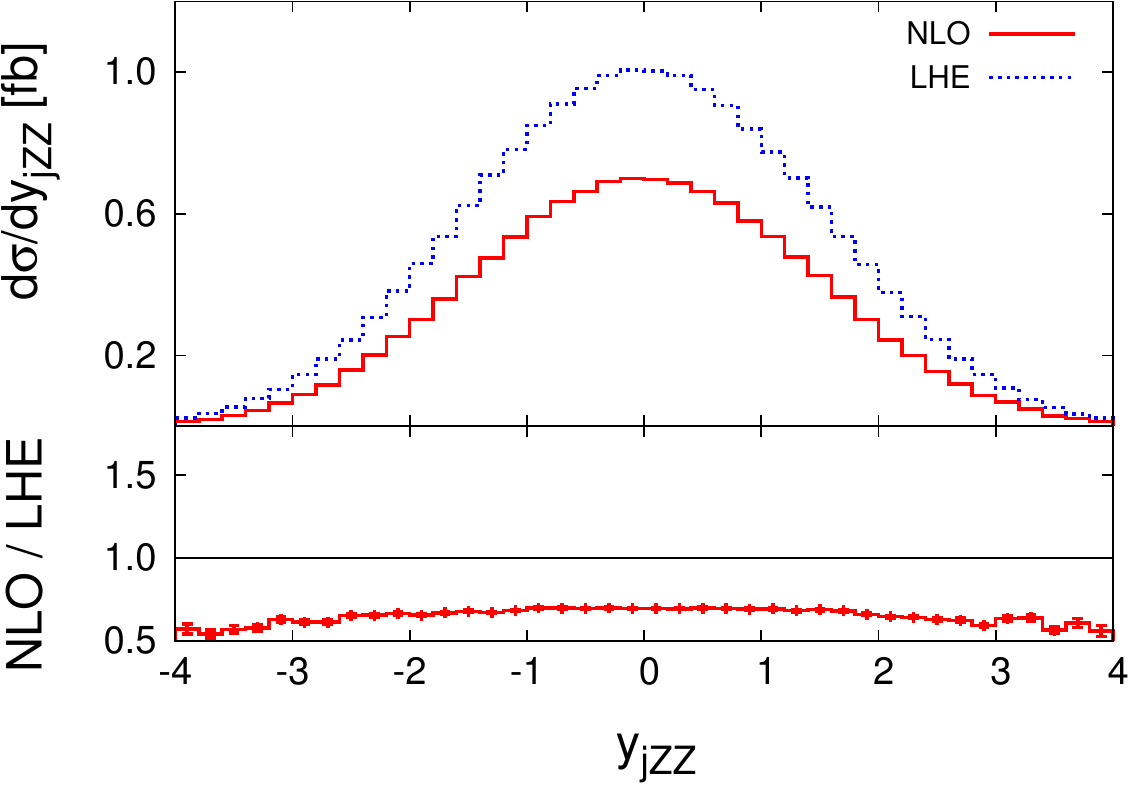} 
\includegraphics[width=0.48\textwidth]{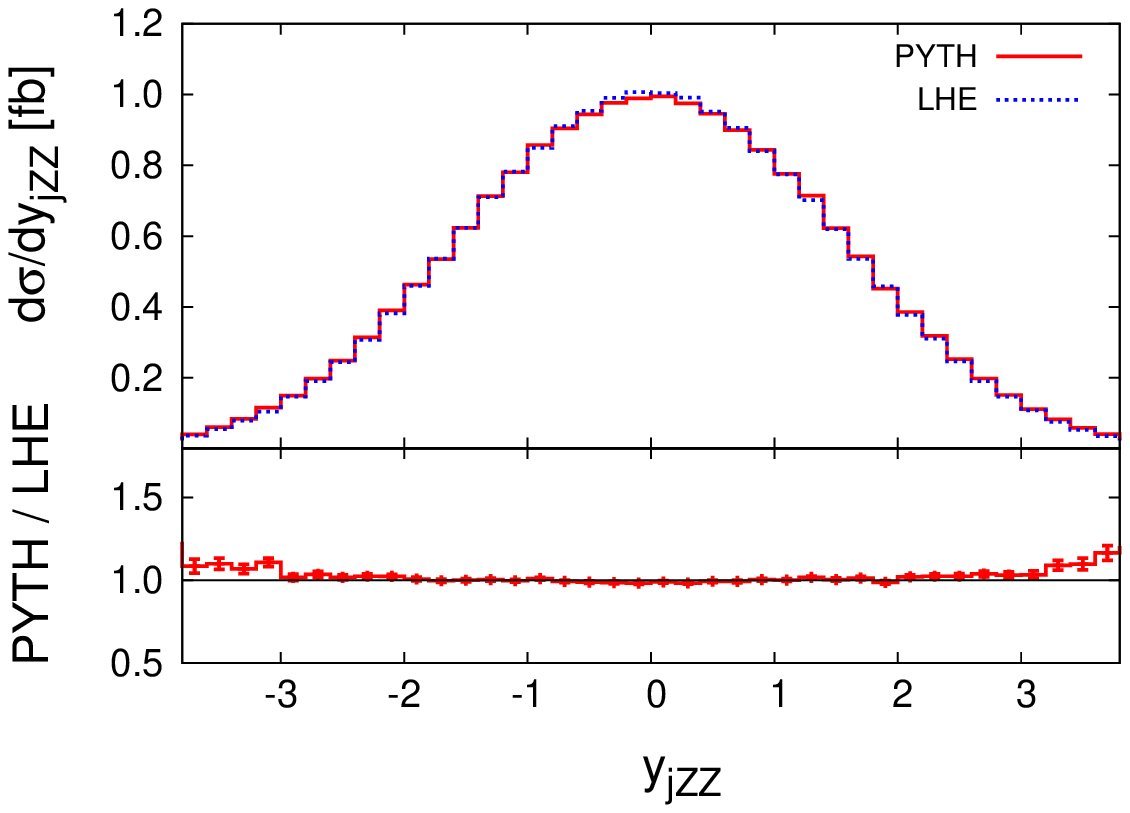} 
\caption{The pseudorapidity of
  the jet $\eta_j$ (top row) and the rapidity difference between the
  jet and the $ZZ$ system, $y_{j\sZ\sZ}$ (bottom row), for $pp\to e^+e^-
  \mu^+\mu^-$ at 7 TeV. Additional cuts of $p_{t, j}>20\;$GeV and
  $\vline \eta_j \vline < 3.5$ are applied. The first column shows
  a comparison between NLO and LHE, and the second column a comparison
  between the pure LHE and the events showered with PYTHIA.}
\label{fig:excl}
} 
Turning now to exclusive variables, we show in Fig.~\ref{fig:excl} the
distribution of the jet pseudorapidity $\eta_j$, and the rapidity
difference between the jet and the $ZZ$ system, $y_{j\sZ\sZ}$, comparing
LHE with both NLO and the effects of showering with \PYTHIA{}.  In
addition to the lepton cuts already described we have jet cuts of
$p_{t, j}>20\;$GeV, and $\vline \eta_j \vline < 3.5$. We see a
noticeable difference between NLO and LHE results in this
distribution. This is a consequence of the jet $\pt$-cut and of the
difference in the jet $\pt$ distribution at NLO and LHE level (see
Fig.~\ref{fig:ptzz}). However, a subsequent parton shower has a small
effect on the LHE results.

\subsection{$W^+W^-/WZ$ Production}

In this Section we present predictions for $W^+W^-/WZ$ production decaying
to four leptons. As for $ZZ$ production, the NLO calculation includes
$Z/\gamma^*$ interference, single resonant diagrams, and when
relevant, interference due to identical leptons.

\begin{table} 
\begin{center}
\begin{tabular}{|c|c|c|c|}
\hline
        & MSTW2008               & CT10                    & NNPDF2.1 \\
\hline
LO (fb) &$18.53(3)^{+0.09}_{-0.25}$&$17.95(3)^{+0.05}_{-0.24}$&$18.30(3)^{+0.15}_{-0.24}$\\
\hline
NLO (fb)&$27.03(4)^{+0.94}_{-0.78}$&$26.00(3)^{+0.91}_{-0.65}$& $26.73(4)^{+0.88}_{-0.73}$\\
\hline
\end{tabular}
\end{center}
\caption{
  As in table \ref{tab:xs} for $pp \to W^-Z \to \mu^- \bar{\nu}_\mu
  e^+ e^- $, with the only cut $M_{e^+e^-}>20$ GeV on the lepton pair
  originating from the $Z$ boson.}
\label{tab:WZ}
\end{table}

\begin{table} 
\begin{center}
\begin{tabular}{|c|c|c|c|}
\hline
        & MSTW2008              & CT10                  & NNPDF2.1 \\
\hline
LO (fb) &$373.8(1)^{+1.2}_{-3.4}$& $368.7(1)^{+1.3}_{-3.4}$&$375.2(1)^{+1.6}_{-3.8}$\\
\hline
NLO (fb)&$498.7(2)^{+13}_{-10}$  & $490.0(2)^{+12}_{-10}$& $499.8(2)^{+12}_{-10}$\\
\hline
\end{tabular}
\end{center}
\caption{
As in table~\ref{tab:xs} for $pp \to W^+W^- \to e^+ \nu_e \mu^-
  \bar{\nu}_{\mu}$, with no cuts applied.}
\label{tab:W^+W^-}
\end{table}

In Tables \ref{tab:WZ} and \ref{tab:W^+W^-}, we show the
cross-sections for $WZ \to e^+e^- \mu^- \bar{\nu}_\mu$ and $W^+W^- \to
\mu^- \nu_{\mu} e^+ \bar{\nu}_e$, for MSTW2008, CT10 and NN2.1 NLO PDF
sets.  The theoretical error due to scale variation is computed with
the same method of table~\ref{tab:xs}.  For $WZ$ and $W^+W^-$
production, we find scale uncertainties of around 1\% at leading
order, and 2-3\% at next-to-leading order. Again the NLO result is not
contained in the LO scale variation band.  This pattern was already
observed and discussed in the $ZZ$ production case.

Our implementation also allows one to study effects due to non
Standard-Model like (anomalous) $WWZ$ and $WW\gamma$ couplings. While
the LHC will probe new physics at the TeV scale directly, anomalous
trilinear gauge couplings (ATGCs) indirectly probe physics at larger scales,
since they arise when high-energy degrees of freedom are integrated
out. Both the Tevatron
\cite{:2009us,Abazov:2009tr,Abazov:2009ys} and LEP
\cite{Alcaraz:2006mx} were able to place quite stringent bounds on
anomalous trilinear couplings.
However, since their effects are enhanced at high
energies, one may expect even better bounds from the LHC. Indeed, CMS
already presented bounds on the anomalous couplings appearing in
an effective Lagrangian with the parametrization of ref.~\cite{Hagiwara:1993ck}
(HISZ from now on) without form
factors \cite{Chatrchyan:2011tz}. It is then interesting to understand
when the LHC running at 7 TeV will be in a position to improve on
existing Tevatron and LEP bounds \cite{Petriello:2011}.

Following ref.~\cite{Hagiwara:1986vm,Baur:1987mt,Hagiwara:1993ck}, we
parametrize the most general terms for the $WWV$ vertex ($V=\gamma,Z$)
in a Lagrangian that conserves $C$ and $P$ as
\newcommand\gwwv{g_{{\scriptscriptstyle WWV}}}
\newcommand\gwwz{g_{{\scriptscriptstyle WWZ}}}
\newcommand\gwwg{g_{{\scriptscriptstyle WW}\gamma}}
\begin{equation}
{\cal L}_{\rm eff} = i \gwwv \left(g_1^V 
(W^*_{\mu\nu} W^\mu V^\nu - W_{\mu\nu}W^{*\mu} V^\nu)
+\kappa^V W^*_{\mu}W_{\nu} V^{\mu\nu}
+\frac{\lambda^V}{M_\sW^2}W^*_{\mu\nu}W^{\nu}_{\rho} V^{\rho\mu}
\right)\,,
\label{leff}
\end{equation}
where $W_{\mu\nu} = \partial_\mu W_\nu - \partial_\nu W_\mu$,
$\gwwz= -e\, {\rm cot} \theta_\sW$ and $\gwwg = -e$.  In the SM
$g_1^V = \kappa_1^V =1$, and $\lambda^V= 0$. Any departure from these
values would be a sign of new physics. Therefore anomalous couplings
are usually written in terms of the deviation from their SM value,
e.g. $\Delta g_1^V = g_1^V-1$ etc.
In our \POWHEG{} generator,
all six parameters can be set independently.
If one imposes gauge invariance under abelian
(electromagnetic) gauge transformations, the parameter $\Delta
g_1^\gamma$ vanishes. This still leaves five independent anomalous
couplings. The LEP groups use a parametrization in which the number of
independent couplings reduces to three.  This is a consequence of only
including operators of up to dimension six in their effective
Lagrangian.  One can write the couplings in terms of the three\newcommand\alphaWphi{{\alpha_{{\scriptscriptstyle W}\phi}}}
\newcommand\alphaBphi{{\alpha_{{\scriptscriptstyle B}\phi}}}parameters $\alpha_\sW, \alphaWphi$ and $\alphaBphi$
\begin{equation}
\Delta g_1^\sZ = \frac{\alphaWphi}{\cos^2\theta_\sW}\,,\quad \lambda^\gamma
= \lambda^\sZ = \alpha_\sW,\quad
\Delta \kappa^\gamma = \alphaWphi+\alphaBphi\,, \quad
\Delta\kappa^\sZ = \alphaWphi-\tan^2\theta_\sW \alphaBphi\,.
\end{equation}
Note that this implies the relation 
\begin{equation}
\Delta \kappa^\sZ = \Delta g_1^\sZ - \Delta \kappa^{\gamma} \tan^2\theta_\sW.
\label{lep}
\end{equation}
In the HISZ model \cite{Hagiwara:1993ck}, it was also suggested that one may set 
$\alphaWphi = \alphaBphi$ as a further simplification, leaving only two independent
parameters. This modified setup is sometimes used in experimental searches.

\FIGURE[t]{ 
\includegraphics[width=0.48\textwidth]{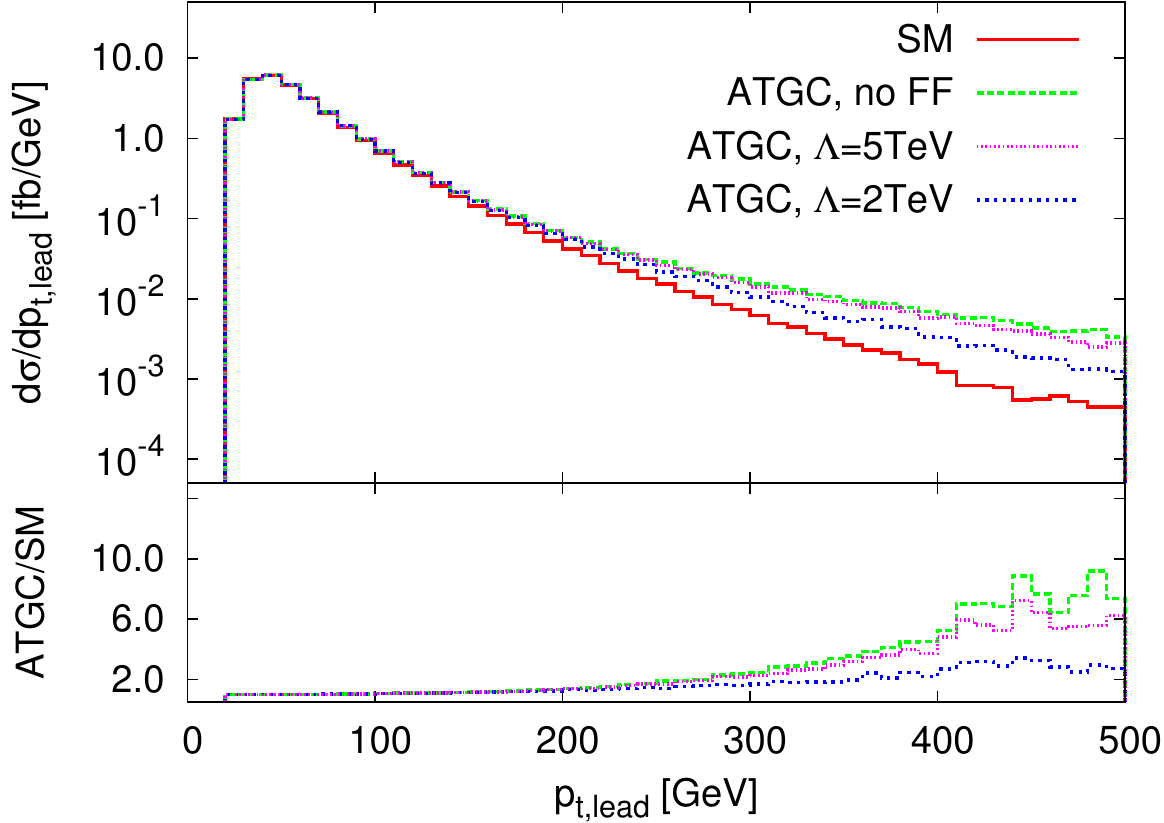}
\includegraphics[width=0.48\textwidth]{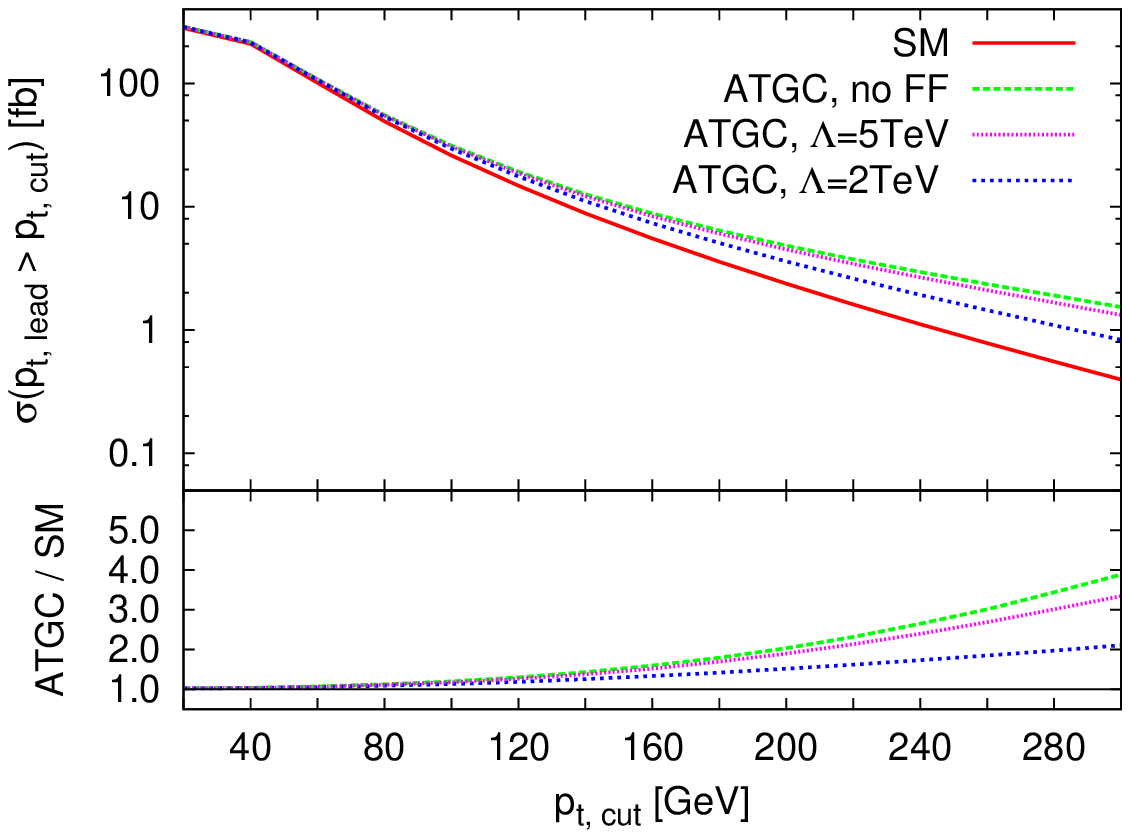}
\caption{The $\pt$ distribution for the leading lepton in
  $W^+W^-$-production is shown on the left, and the integrated cross-section
  as a function of the cut on the transverse momentum of the leading
  lepton is shown on the right. The results are shown using Standard
  Model couplings (in red, labeled `SM'), as well as with the
  anomalous couplings of eq. \eqref{atgcs} (in green, labeled `ATGC,
  no FF'). The effect of a form factor given in eq. \eqref{ff} is also
  shown, both for a value of $\Lambda=5$~TeV and $\Lambda=2$~TeV (in
  pink and blue respectively).}
\label{fig:ptlead}} 

In the presence of anomalous couplings, the effective Lagrangian of
eq.~\eqref{leff} gives rise to interactions that violate unitarity at
high energy. Thus, in order to achieve a more realistic parametrization,
the couplings are multiplied by
form factors, that embody the effects arising from integrating out the
new physics degrees of freedom. The precise details of the form factors
therefore depend on the particular model considered. Paralleling here
the discussion of ref.~\cite{Baur:1988qt},
we assume that all anomalous coupling $\Delta
g$ are modified as
\begin{equation} 
\Delta g \to \frac{\Delta g}{(1+ M_{\sV\sV}^2/\Lambda^2)^2}\,,
\label{ff}
\end{equation}
where $M_{\sV\sV}$ is the invariant mass of the vector boson pair and
$\Lambda$ is the scale of the new physics.

In the LEP study of ref.~\cite{LEP:2002mc} the value of one or two of
the parameter(s) is fixed to their SM values, 
and determines the range of values of the remaining parameter(s) consistent with their data. 
We choose the maximum deviation of \textit{all} parameters from their SM values allowed by these fits. 
While it is true that this point in parameter space would lie outside the LEP bounds where all three parameters are allowed to vary, we choose it for illustrative purposes.

 The values of the parameters are
\begin{equation}
\Delta g^\sZ  = -0.027, \qquad \Delta \lambda^\sZ
 =  \Delta \lambda^{\gamma}  = -0.044, \qquad 
\Delta \kappa^{\gamma} = -0.112,
\label{atgcs}
\end{equation}
and where (using the LEP groups parametrization) $\Delta \kappa^\sZ$ is
obtained from the above using eq.~\eqref{lep}. We note that these
values for the ATGCs also fall within the unitarity bounds of
ref.~\cite{Aihara:1995iq} for $\Lambda \lesssim 7$ TeV.
In the left-hand plot of Fig.~\ref{fig:ptlead} we
show the $\pt$ distribution for the hardest lepton (which is known to be very sensitive to ATGCS) in $W^+W^-$-production
using the above ATGCs, with
different values of the form factor parameter ($\Lambda = 2$ TeV, 5 TeV and
$\infty$).
The effect of the anomalous couplings is evident in the high-$\pt$
tail of the distribution.
The form factor damps the effect of the ATGCs. Its impact increases
with decreasing $\Lambda$, while $\Lambda=\infty$
corresponds to no form factor. In the
right-hand plot of Fig.~\ref{fig:ptlead}, we show the integrated
cross-section as a function of the cut on the $\pt$ of the leading lepton.
Fig.~\ref{fig:ptlead} gives an indication of the kind of signal one
would see at the LHC. Depending upon the precise form of the new physics,
the signal would lie somewhere between the
$\Lambda = 2$ TeV and $\Lambda = \infty$ predictions.
We note that in the region $\pt >
p_{t,\mathrm{min}}$, with $150 \lesssim p_{t,\mathrm{min}} \lesssim
250$ GeV, the cross section is strongly affected by the anomalous couplings.
Furthermore, its value is of the order of 1 to 10 fb,
and thus such a difference should be observable with a few
inverse femtobarns of data. It may then be possible for the LHC
to improve on the LEP bounds on anomalous trilinear boson couplings
already by the end of the year.

\FIGURE[t]{ 
\includegraphics[width=0.48\textwidth]{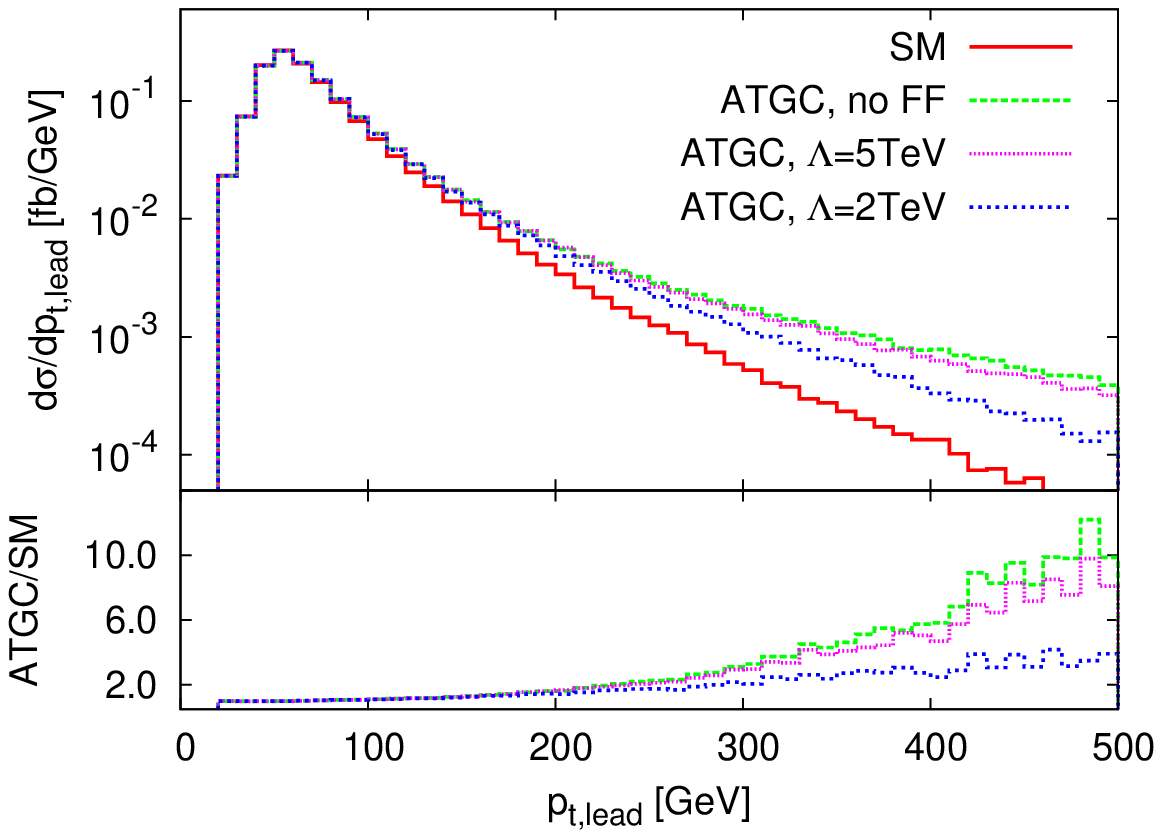}
\includegraphics[width=0.48\textwidth]{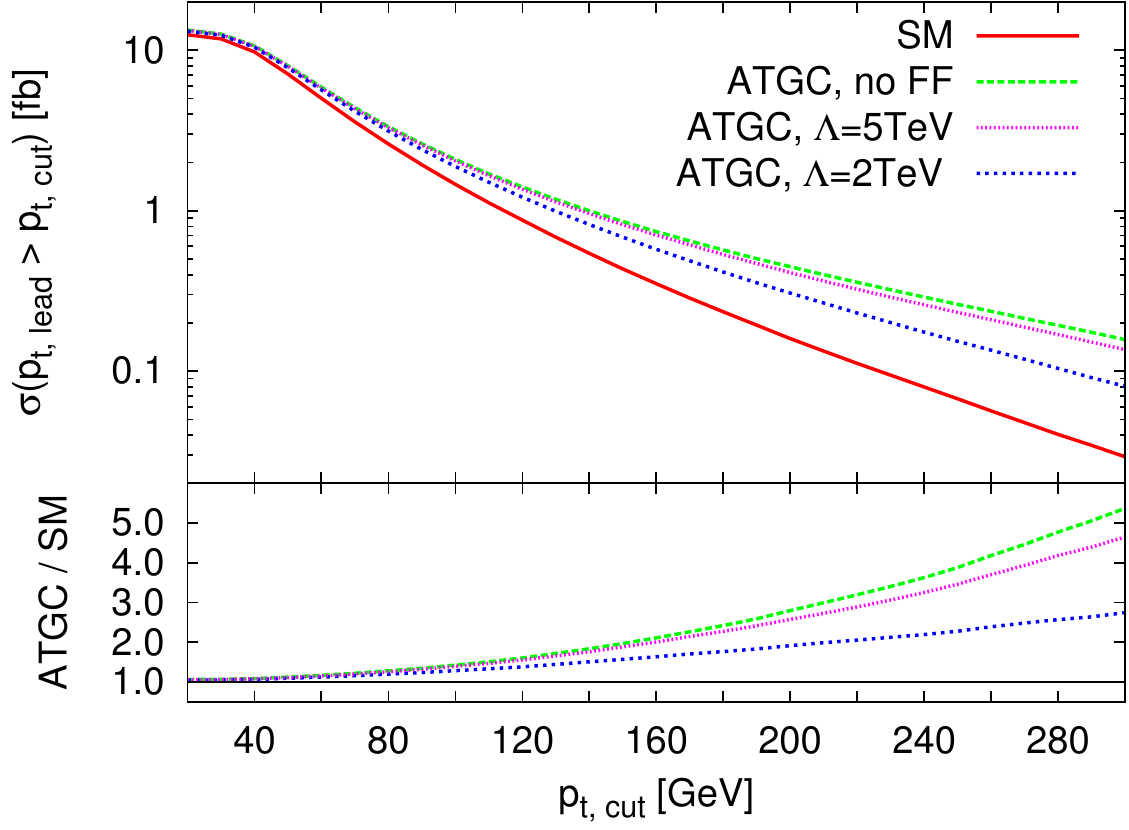}
\caption{As in Fig.~\ref{fig:ptlead} for $W^-Z$ production.}
\label{fig:ptleadWZ}} 

Similar results are obtained in the case of $WZ$ production, and are
shown in Fig.~\ref{fig:ptleadWZ} for $W^-Z$.  This process is
sensitive only to the $WWZ$ tri-linear coupling and not to the
$WW\gamma$ couplings.  In this case the cross section is smaller and
one expects to make less stringent bounds than in the $W^+W^-$ study.

\section{Conclusion}
\label{sec:conclu}
We have implemented the production of vector boson pairs $W^+W^-$,
$W^{\pm}Z$ and $ZZ$ in the \POWHEGBOX. Off-shell effects, $Z/
\gamma^*$ interference, single resonant amplitudes and interference
between same-flavour leptons have all been included. The only effect
not included is the interference between $W^+W^-$ and $ZZ$ processes,
when both decay to same-flavour leptons and neutrinos. We have,
however, shown that this interference is entirely negligible, so that
it suffices to consider these processes separately. We have shown
that, for variables inclusive in the hardest jet (or equivalently
in the radiated parton),
there is very little difference between
NLO and LHE results, whereas noticeable differences arise
for variables that probe the hardest jet.
Neither kind of observable is much affected by the subsequent parton shower.

Our \POWHEG{} implementation allows the study of the effect of
anomalous trilinear boson couplings in $WZ$ and $W^+W^-$
production. We demonstrated this for both processes at the LHC with
$\sqrt{s}=7$ TeV, and looked at the distribution of the $\pt$ of the
hardest lepton. We saw that a set of anomalous couplings allowed by
the LEP bounds leads to a large deviation from the Standard Model
result in the high $\pt$ bins of this distribution, to such an extent
that the LHC may be able to probe this by the end of the present year.

We have not included in this work the $W\gamma$ or $Z\gamma$ processes
with an on-shell photon.  This is because NLO calculations with final
state on shell photons have to dealt with in a different way.  Photons
can be radiated in the collinear direction by light quarks, and thus
generate singularities that require special attention. Including
photons in a \POWHEG{} generator is however possible, and in
ref.~\cite{D'Errico:2011sd}, for example, the $\gamma\gamma$
production process was implemented. We thus defer this problem to
future work.

Finally, the code for our generators has been made
available at the \POWHEGBOX{} website
\url{http://powhegbox.mib.infn.it}.

\vspace{1cm} {\bf \noindent Acknowledgments} 
We wish to thank J. Campbell, S. Dittmaier, K. Ellis, F. Maltoni,
R. Frederix and G. Passarino for useful exchanges.  This work is
supported by the British Science and Technology Facilities Council, by
the LHCPhenoNet network under the Grant Agreement PITN-GA-2010-264564
and by the European Research and Training Network (RTN) grant
Unification in the LHC ERA under the Agreement PITN-GA-2009-237920.

\appendix 
\section{Phase space sampling}
Although the \POWHEG{} implementation of the processes we consider is
relatively straightforward, some care is needed in the generation of
the Born phase space, to ensure that the resonant regions are probed
efficiently. Thus, when implementing processes with no lepton
interference, like $ZZ\to e^+ e^- \mu^+ \mu^-$, the phase space
generator should include the virtuality of the $e^+ e^-$ and $\mu^+
\mu^-$ pairs as integration variables, so that the importance sampling
can be performed near the $Z$ peak or for low invariant mass, where
the pole of the photon propagator starts to count.  In $ZZ$ and $WZ$
production, a cut on the lepton pair mass for oppositely charged same
flavour leptons must be imposed, in order to stay away from the photon
pole.

When lepton interference is present, like in $ZZ\to e^+ e^- e^+ e^-$,
there are two ways to assign the leptons to the vector resonance,
and using all the corresponding
mass combinations as integration variables becomes too
cumbersome. In
this case, in order to maintain an efficient importance sampling,
we do the following.  Calling 1 and 2 the two possible
resonance assignments, we multiply the cross section by a factor $2
F_2/(F_1+F_2)$, where the $F_{1/2}$ are factors that suppress the
vector poles for the 1 or 2 resonance assignment. This is possible
since the two resonance assignments differ only by a permutation of
identical fermions, and the cross section is symmetric under
$1\leftrightarrow 2$ exchange.  After multiplying the cross section by
this factor, the poles of the second region are suppressed, and one
can adopt a phase space importance sampling as if there was only the
resonance assignment 1. In case of the $ZZ\to e^+(1) e^-(2) e^+(3)
e^-(4)$ process, for example, the function $F_1$, relative to the
resonance assignment $(1,2),(3,4)$ is chosen equal to
\begin{equation}
F_1=\left\{ s_{12}\left[(s_{12}-M_\sZ^2)^2+\Gamma_\sZ^2 M_\sZ^2\right]\times
       s_{34}\left[(s_{34}-M_\sZ^2)^2+\Gamma_\sZ^2 M_\sZ^2\right]\right\}^2,
\end{equation}
while $F_2$ is obtained by replacing $1 \leftrightarrow 3$.
In case of the $ZW \to e^+(1)e^-(2) e^+(3) \nu_e(4)$ process, we have
\begin{equation}
F_1=\left\{ s_{12}\left[(s_{12}-M_\sZ^2)^2+\Gamma_\sZ^2 M_\sZ^2\right]\times
       \left[(s_{34}-M_\sW^2)^2+\Gamma_\sW^2 M_\sW^2\right]\right\}^2,
\end{equation}
for the region $(1,2),(3,4)$. For $F_2$ one replaces again $1
\leftrightarrow 3$.  In order to avoid possible errors in the
analysis, the output at the level of Les Houches Interface is
symmetrized again by randomly permuting the kinematics of the
identical final state leptons.

\section{Born zeros}
All three production processes considered here have a Born cross
section that vanishes in particular kinematic regions due to angular
momentum conservation. In fact, since the vector interaction preserves
chirality, the incoming quark-antiquark system has total spin equal to
1 in magnitude.  If we consider the final state configuration where
all leptons are parallel to the incoming quark, the whole system is
symmetric under azimuthal rotations, and thus the angular momentum
component along the collision axis equals the sum of the spins, and
should be preserved.  On the other hand, the final state leptons can
be divided into two pairs with opposite helicity, thus yielding two
systems with total spin 1 (in absolute value) along the collision
axis. Since the sum of the two spin 1 system yields either 2 or zero,
angular momentum is violated in this configuration, and the amplitude
must vanish. This property can be easily verified numerically.

It was first pointed out in ref.~\cite{Alioli:2008gx} that problems
can arise in the \POWHEG{} formalism if the Born cross section
vanishes or becomes particularly small in certain kinematic
regions. We briefly recall the nature of this problem in the
following.

The basic \POWHEG{} cross section formula can be written as
\begin{equation}\label{eq:pwghardest}
d\sigma =\bar{B}(\Phi_B)d\Phi_B\left[\Delta^s_{t_0}(\Phi_B)+\Delta^s_t(\Phi_B)
\frac{R^s(\Phi)}{B(\Phi_B)} d\Phi_r\right]+[R(\Phi)-R^s(\Phi)]d\Phi,
\end{equation}
where the real phase space is factorized in terms of the underlying
Born $\Phi_B$ and the radiation $\Phi_r$ phase space: $d\Phi= d\Phi_r
d\Phi_B$, and $B$, $R$ and $V$ are the Born, real and virtual
amplitudes. $R^s$ is an approximation to the real amplitude such that
$0\le R^s\le R$, and that $R^s\to R$ in the limit of soft or collinear
singularities.  The choice $R^s=R$ is also allowed, and is often used.
The variable $t$ represents, in the present case, the transverse
momentum of the radiated parton relative to the collision axis
(thus $t$ is a function of the phase space point),
and $t_0$ is a non-perturbative cutoff (of the order of a typical
hadronic scale) on this variable.
Furthermore
\begin{equation}\label{eq:bbar}
\bar{B}(\Phi_B)=B(\Phi_B)+\left[V(\Phi_B)+\int R^s(\Phi) d\Phi_r\right],
\end{equation}
and
\begin{equation}
\Delta^{\rm s}_{t_l}(\Phi_B)=\exp\left[-\int_{t>t_l} \frac{R^{\rm s}(\Phi)}{B(\Phi_B)} d\Phi_r
\right]\;.
\end{equation}
The Born term may become
particularly small in certain kinematic regions; for example, if it
vanishes for symmetry reasons. Besides the present case, another
example of a behaviour of this kind is the case of $W$ production and
decay, where there is a zero in the Born cross section if the outgoing
lepton is anti-parallel to the incoming quark \cite{Alioli:2008gx},
which is due to angular momentum conservation and to the left-handed
nature of charged currents. It may then turn out that the $\bar{B}$
function does not vanish in the same region, due to the presence of
the real term in eq.~\eqref{eq:bbar}. From eq.~\eqref{eq:pwghardest},
we see that, in this case, away from the Sudakov region, the real
contribution may be enhanced by a factor $\bar{B}/{B}$. The
\POWHEGBOX{} has a built-in mechanism to deal with this problem. If
the flag {\tt withdamp} is set, $R^s$ is chosen to vanish in the
regions where $R$ differs too much (by more than a factor of 5) from
its collinear or soft approximation, which are proportional to the
underlying Born cross section. The contribution $R-R^s$, being
non-singular, is then added independently. We thus conclude that the
{\tt withdamp} flag should also be activated in the processes we are
considering.

In practice, it turns out that the effect of the {\tt withdamp} flag
is much more important in the $W^+W^-$ and in the $WZ$ case, than in
the $ZZ$ case. This fact can be easily understood in the $WZ$ case,
where another region of vanishing Born cross section can be found
which is less suppressed by phase space. Assume we have an incoming
$d$ quark in the positive rapidity direction, colliding with a
$\bar{u}$. In $WZ$ production only the left handed component of the
incoming $d$, and the right handed component of the incoming $\bar{u}$,
contribute, so that the incoming particles have a definite angular
momentum projection along the collision axis equal to 1. Consider now
the kinematic region where the $W^-$ decay products are aligned along
the collision axis, with the negative lepton in the negative rapidity
direction. Again, the negative lepton is left handed, while the
antineutrino is right handed.  Thus the system has angular momentum
projection along the collision axis equal to -1. Since the $Z$ must
also have zero transverse momentum, in order to balance the angular
momentum component along the collision axis it should have spin 2,
which is impossible. Thus this region is also suppressed, irrespective
of the direction of the decay products of the $Z$.

The $W^+W^-$ case is not as simple, since here there are two
amplitudes that contribute: one with the two $W$s emerging from the
incoming fermion line (t-channel amplitude), and the other with the
two $W$s arising from the $WWZ$ or $WW\gamma$ vertex, the $Z/\gamma^*$
being attached to the incoming quark line (s-channel amplitude).  The
same argument applied to the $WZ$ case also applies to the $t$-channel
amplitude. We can use a similar argument to show that the $s$-channel
amplitude vanishes when the decay products of the two $W$'s travel
along the same line, with the neutrino of one $W$ is aligned to the
lepton of the other. In this case, the decay system has angular
momentum projection in the line of decay equal to 2, and it can't
therefore arise from a virtual $Z$ or $\gamma$ decay. Although we
cannot conclude that the full amplitude vanishes in either direction,
even here we expect a strong kinematic suppression of the Born
amplitude in certain regions.

\FIGURE[t]{
\includegraphics[width=0.48\textwidth]{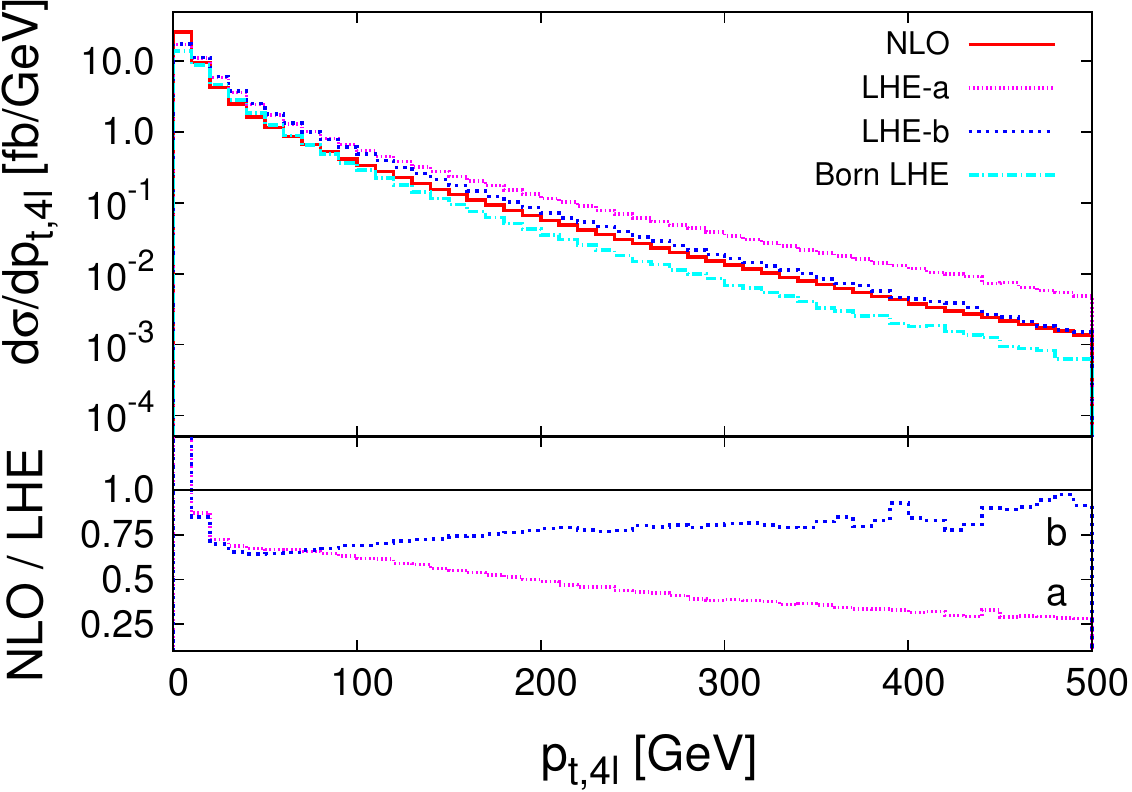}
\includegraphics[width=0.48\textwidth]{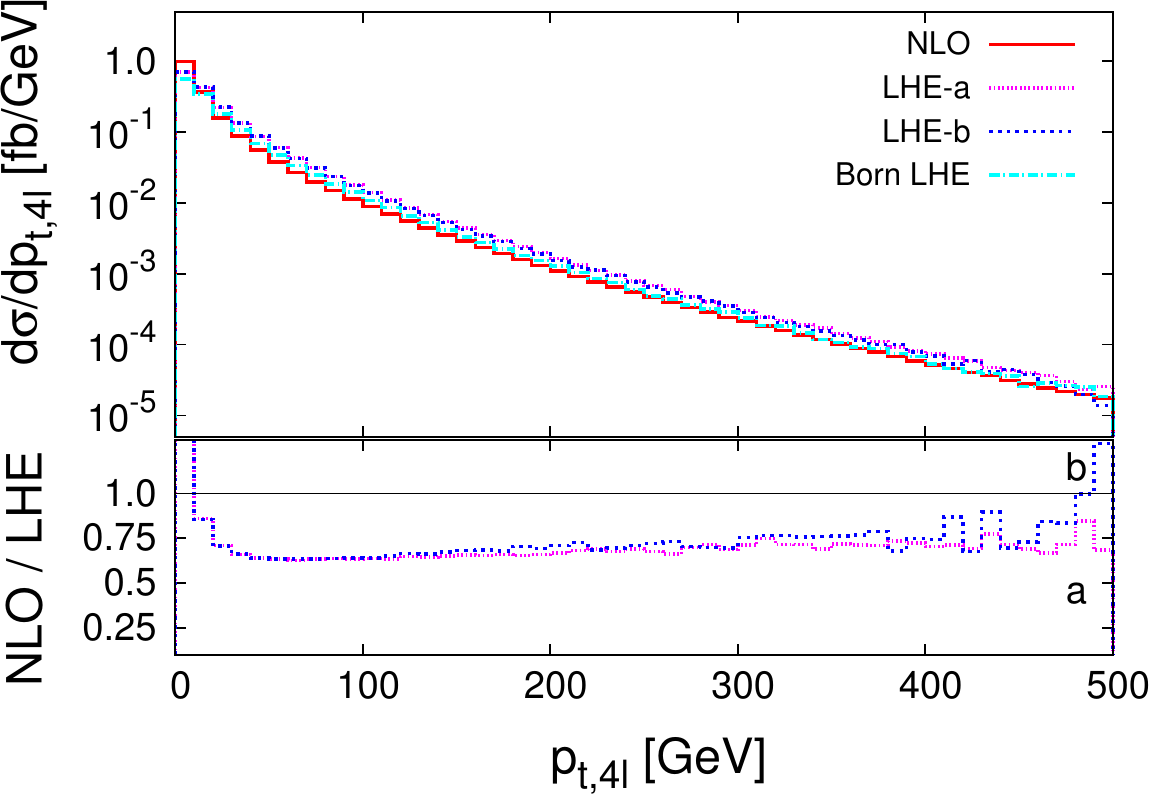}
\caption{Distribution of $p_{t,4l}$ for $pp \to W^+W^- \to e^+ \nu_e \mu^-
  \bar{\nu}_{\mu}$ and $pp \to ZZ \to e^+ e^- \mu^+ \mu^-$, at centre
  of mass energy 7 TeV. The NLO results are shown in red, and LHE
  results using only the Born cross-section to generate events are
  shown in light blue. Also shown are the NLO LHE results with damping
  (in dark blue and labeled `LHE-b') and without damping (in magenta
  and labeled `LHE-a'). The bottom plot shows the ratio of the NLO
  results to the LHE results with damping (dark blue, labeled `b'),
  and the ratio of the NLO results to the LHE results without damping
  (magenta, labeled `a').  }
\label{fig:bornzero}}

In Fig.~\ref{fig:bornzero}, we show the transverse momentum
distribution of the four lepton system in $ZZ$ and $W^+W^-$ production
if the {\tt withdamp} flag is not set \footnote{Because of the
  radiation zeroes, the upper bound of radiation tends to diverge as
  we increase the number of points used to compute it (i.e. the {\tt
    nubound} variable in the {\tt powheg.input} file).  On the other
  hand, with a limited number of points we still get a relatively
  small rate of upper bound violations in the generation of radiation,
  so that these distributions can be nevertheless computed.} (the
effect in $WZ$ production is very similar to that in $W^+W^-$
production and is not shown).  While we find a very modest effect in
the $ZZ$ case, in the $W^+W^-$ case we find large enhancement of the
differential cross section as a function of the LHE transverse
momentum of the hardest parton. We also find that this anomalous
behaviour disappears if we set $\bar{B}=B$ in \POWHEG{} (this is
achieved by setting the flag {\tt bornonly} to 1 in the {\tt
  powheg.input} file).  Thus, the anomalous growth is due to a large
$\bar{B}/B$ ratio in some kinematical regions.

\end{document}